\documentclass[aps,prb,floatfix,amsmath,amssymb,preprint,eqsecnum,nofootinbib,superscriptaddress]{revtex4}
\usepackage{graphicx}
\usepackage{dcolumn}
\usepackage{bm}
\usepackage{epstopdf}
\usepackage{setspace}   

\newcommand{\beq}{\begin{equation}}
\newcommand{\eeq}{\end{equation}}
\newcommand{\ba}{\begin{array}{ccc}}
\newcommand{\ea}{\end{array}}
\newcommand{\nn}{\nonumber}
 \renewcommand{\d}{\partial}
\def\bea{\begin{eqnarray}}
\def\eea{\end{eqnarray}}

\def\<{\langle}
\def\>{\rangle}

\usepackage{amsmath}
\usepackage{amssymb}

\begin{document}

\title{A symmetry-respecting topologically-ordered surface phase of 3d electron topological insulators}

\author{Max A. Metlitski}
\affiliation{Kavli Institute for Theoretical Physics, UC Santa Barbara, CA 93106}
\author{C. L. Kane}
\affiliation{Department of Physics and Astronomy, University of Pennsylvania, Philadelphia, PA 19104}
\author{Matthew P. A. Fisher}
\affiliation{Department of Physics, UC Santa Barbara, CA  93106}

\date{\today\\
\vspace{1in}}
\begin{abstract}
A 3d electron topological insulator (ETI) is a phase of matter protected by particle-number conservation and time-reversal symmetry. It was previously believed that 
the surface of an ETI must be gapless unless one of these symmetries is broken. A well-known symmetry-preserving, gapless surface termination
of an ETI supports an odd number of Dirac cones.  In this paper we deduce a symmetry-respecting, {\it gapped} surface termination of an ETI, which carries an 
intrinsic 2d topological order, Moore-Read$\times U(1)_{-2}$. The Moore-Read sector supports non-Abelian charge $1/4$ anyons, while
the Abelian, $U(1)_{-2}$, (anti-semion) sector is electrically neutral. Time-reversal symmetry is implemented in this surface phase in a highly non-trivial way. 
Moreover, it is impossible to realize this phase strictly in 2d,  simultaneously preserving its implementation of both the particle number and time-reversal symmetries.
A 1d edge on the ETI surface between the topologically-ordered phase and the topologically trivial time-reversal-broken phase with a Hall conductivity $\sigma_{xy} = 1/2$, carries a right-moving
neutral Majorana mode, a right-moving bosonic charge mode and a left-moving bosonic neutral mode.  The topologically-ordered phase is separated from the surface superconductor 
by a direct second order phase transition in the $XY^*$ universality class, which is driven by the condensation of a charge $1/2$ boson, when approached from the topologically-ordered side, and proliferation of a flux $4 \pi$ vortex, when 
approached from the superconducting side. In addition, we prove that time-reversal invariant (interacting) electron insulators with no intrinsic topological order
and electromagnetic response characterized by a $\theta$-angle, $\theta = \pi$, do not exist if the electrons transform as Kramers singlets under time-reversal.
\end{abstract}

\maketitle

\section{Introduction.}
The past few years have seen impressive advances in our understanding of quantum phases of matter endowed with
a global symmetry. Particularly remarkable progress has been made in the study of so-called symmetry
protected topological (SPT) phases.\cite{TI,FuKaneMele, MooreBalents, Roy, Hasan, Ludwig, KitaevFree, Fidkowski1d, Ari, Wen1d, Cirac, Wen1dfull, WenCohoBoson, Ashvin2d, VS} SPT phases are fully gapped states of matter which possess a global symmetry. 
When this symmetry is present, an SPT phase cannot be connected to a trivial product state without a phase
transition. On the other hand, once the symmetry is broken, an SPT state may be continuously deformed into a trivial product state.
Therefore, SPT phases carry no intrinsic topological order: they have no ground state degeneracy on a torus, support no bulk excitations with 
fractional statistics and quantum numbers, and possess no long-range entanglement.

Thus, viewed from the perspective of bulk excitations, SPT phases are trivial. However, they possess highly unusual 
boundary states. In fact, the 0d boundary of a 1d SPT phase is always gapless, the 1d edge of a 2d SPT phase is either gapless or spontaneously breaks
the symmetry, and the 2d surface of a 3d SPT is either gapless, spontaneously breaks the symmetry or carries an ${\it intrinsic}$ 2d topological order.
Moreover, in all cases, the symmetry on the boundary of a $d$-dimenional SPT phase is realized in a way that is impossible in a strictly  $(d-1)$-dimensional system.
The last fact makes SPT phases not only interesting in their own right, but also places them in the context of a very broad question: ``What are the consistency conditions
on symmetry implementation?" 

The fact that a gapped symmetry-preserving termination of a 3d SPT phase may exist if one allows for the possibility of intrinsic topological order on the surface has not 
been recognized until the recent work of A.~Vishwanath and T.~Senthil.\cite{VS} Ref.~\onlinecite{VS} proposed SPT phases of interacting bosons in 3d with several symmetries 
and deduced their gapped, symmetry-respecting topologically-ordered surfaces. To specify these surface phases one must identify both the intrinsic topological order (anyon types, fusion rules and braiding statistics),
as well as the transformations of the anyons under the global symmetry. For instance, Ref.~\onlinecite{VS} proposed a 3d SPT phase of bosons protected by the symmetry, $U(1)\ltimes {\cal T}$, with $U(1)$ - the particle-number conservation and ${\cal T}$ - time reversal.\footnote{The semi-direct product $\ltimes$ means that the anti-unitary time reversal operator ${\cal T}$ and the $U(1)$ rotations $g$ do not commute,
${\cal T}^{-1} g {\cal T} = g^{-1}$.} A symmetry-respecting gapped surface of this phase supports a toric code topological order, whose anyons $e$ and $m$ both carry charge $1/2$ under the $U(1)$ symmetry. Strictly in 2d, such charge assignment would require the presence of a non-zero electrical Hall conductivity, $\sigma_{xy}$, and is, thus, incompatible with time-reversal symmetry.\cite{VS, StatWitt} However, on the surface of the 3d SPT phase, such a toric code has $\sigma_{xy} = 0$ and preserves the full $U(1)\ltimes {\cal T}$ symmetry. An explicit coupled-layer construction of this phase has appeared in Ref.~\onlinecite{Chong}.

Symmetry-respecting topologically-ordered surface states provide a convenient label for their corresponding 3d SPT phases. This should be compared to the case of 2d SPTs, which are naturally labeled by their gapless edge conformal field theory (CFT). Unfortunately, our understanding of CFTs (let alone more general gapless states) in 2d is far less complete than in 1d: as a result,  we are not presently in a position to characterize most 3d SPT phases by their gapless surfaces. In contrast, our understanding of 2d topologically-ordered phases is fairly well-developed. Moreover, much progress has  recently been made in the study of phases with intrinsic topological order that are, in addition, endowed with a global symmetry; such states are refered to as  ``symmetry enriched topological" (SET) phases.\cite{Hermele, Ran, WenSET, Chong, AshvinZ2Z2} By studying topologically-ordered terminations of 3d SPT phases one can learn which SET phases are prohibited strictly in 2d.\cite{Chong} Conversely, given an intrinsic topological order and a global symmetry $G$, one may search for a set of anyon transformation laws under $G$, which is consistent with the fusion and braiding rules, but, nevertheless, cannot be realized strictly in 2d. One may then be able to deduce a 3d SPT phase with the corresponding topological order and symmetry implementation on its surface.\cite{AshvinToricf,LukaszTSC}

Probably the most famous and one of the very few experimentally realized examples of an SPT phase is provided by the 3d electron topological insulator (ETI).\cite{TI, FuKaneMele, MooreBalents, Roy, Hasan} An ETI is a phase of fermions (electrons), protected by particle-number conservation symmetry, $U(1)$, and time-reversal symmetry, ${\cal T}$. An important point is that these two symmetries are linked by the relation, ${\cal T}^2 = (-1)^{N}$, where $N$  is the particle number. This relation is obeyed by the standard transformation law for spinful electrons under time-reversal, ${\cal T}: c_{\alpha} \to \epsilon_{\alpha \beta} c_{\beta}$. Thus, electrons are Kramers doublets. Conveniently, an ETI can be realized even with non-interacting electrons. As long as the interactions are not too strong, an ETI surface respects the $U(1)$ and ${\cal T}$ symmetries and supports an odd number of gapless Dirac cones.\footnote{Strictly speaking, if the chemical potential on the surface is away from the Dirac point, the surface will be in a Fermi-liquid phase, which possesses a superconducting instability for arbitrarily weak attraction in the BCS channel.} It is well known that such properties cannot be realized strictly in 2d.\cite{Redlich, Semenoff, FionaMike, StatWitt} Note that the massless Dirac theory is a rare example of a 2d CFT that is well-understood.  

The fact that the ETI surface cannot be mimicked in a strictly 2d system is intimately related to the electromagnetic response in the ETI bulk, which is characterized by the so-called $\theta$-angle, $\theta = \pi$.\cite{SCZ} The $\theta$-angle is directly manifested in the Witten effect: a magnetic monopole in the ETI bulk carries a half-odd-integer electric charge.\cite{Witten,SCZ,Franz}

If the interactions on the ETI surface are sufficiently strong, the $U(1)$ or ${\cal T}$ symmetry may become spontaneously broken and the Dirac cone will be gapped out. The same effect is achieved by explicitly breaking  the symmetry on the surface. For instance, if one breaks ${\cal T}$ by coating the surface with a magnetic insulator, one obtains a fully gapped $U(1)$-preserving state with no intrinsic topological order and a Hall conductivity, $\sigma_{xy} = \pm 1/2$. Alternatively, one can break the $U(1)$ symmetry, but preserve ${\cal T}$, by coating the surface with an $s$-wave superconductor. The resulting superconducting (SC) surface phase again supports no intrinsic topological order and possesses gapped Bogolioubov quasiparticle excitatations. A curious property of the surface superconductor is that the vortices with flux $\Phi = \pi$ carry Majorana zero modes and have (projective) non-Abelian statistics.\cite{FuKaneMajorana} (We follow the standard convention, where flux $\Phi = 2 \pi$ corresponds to $\Phi = h c/e$). 

Thus, we have a full understanding of both the symmetry-respecting, gapless surface termination of an ETI and the symmetry breaking terminations. Elementary arguments based on the Witten effect in the ETI bulk indicate that if the ETI surface is gapped and symmetry-respecting, it must be topologically-ordered.\cite{StatWitt} One may then wonder if such a gapped symmetry-respecting ETI termination actually exists (at least, theoretically).  In this paper, we will deduce such a surface phase, providing a positive answer to the above question. 

We would like to note that an analogous question was recently addressed by Ref.~\onlinecite{LukaszTSC} in the context of another 3d SPT phase of fermions - the topological superconductor. A topological superconductor is an SPT phase protected solely by time-reversal symmetry, with fermions transforming as Kramers doublets.\cite{KitaevFree,Ludwig, Balian, Volovik} Just as the ETI, this phase can be obtained with non-interacting fermions. Its gapless ${\cal T}$-respecting termination supports a single Majorana cone. 
In a technical masterpiece, the authors of Ref.~\onlinecite{LukaszTSC} use an exactly solvable Walker-Wang model\cite{WW} to construct a 3d SPT phase of fermions protected by time-reversal symmetry, which possesses a ${\cal T}$-respecting gapped topologically-ordered surface. They further conjecture that the SPT phase obtained in this manner is smoothly connected to the well-known non-interacting topological superconductor. In this paper, we will utilize a rather different approach to constructing the topologically-ordered surface phase of an ETI, which is more grounded in the known ETI physics. In particular, our approach leaves little doubt that the topological order obtained can, in principle, be realized on the ETI surface.



\section{Overview.}

We now give a brief overview of our construction and results. We start with the superconducting surface phase of an ETI and imagine ``quantum disordering" it by proliferating vortex defects. We will argue that the smallest vortex that can ``condense'' carries magnetic flux, $\Phi = 4 \pi$. Proliferation of these vortices drives the system into a time-reversal symmetric insulating phase with an intrinsic topological order, Moore-Read$\times U(1)_{-2}$. This topological order is non-Abelian and supports 24 types of anyons, counting the electron (12, if we identify excitations differing by an electron). The $U(1)_{-2} = \{1, \bar{s}\}$ sector is Abelian; its single non-trivial anyon, $\bar{s}$, is an electrically neutral anti-semion. The Moore-Read sector is identical in its intrinsic topological order and anyon electric charges to the famous electron quantum Hall state at $\nu = 1/2$. It can be thought of as a subset of the Ising$\times U(1)_{8}$ topological order consisting of the following anyons
\bea &&\sigma e^{i m \varphi_\rho}, \quad\quad \quad \quad \quad \quad \,\,\,\,\,\,\,\,\,\,\, m = 1, 3, 5, 7,\quad\quad Q = m/4\nn\\
&& e^{i m \varphi_\rho} ,\,\,\,\,\,\,\, fe^{i m \varphi_\rho}, \quad \quad\quad\quad  m = 0, 2, 4, 6,\quad\quad Q = m/4
\label{Eq:MooreRead} \eea
Here, the labels $\{1, \sigma, f\}$ run over the Ising sector and $e^{i m \varphi_\rho}$, $m = 0\ldots 7$, run over the $U(1)_8$ sector. Note that the Ising and $U(1)_8$ sectors have the ${\it same}$ chirality, which is opposite to the chirality of the $U(1)_{-2}$ sector. In eq.~(\ref{Eq:MooreRead}), $Q$ denotes the electric charge of the anyons. 
The non-Abelian anyons $\sigma e^{i m \varphi_\rho}$ and $\sigma e^{i m \varphi_\rho} \bar{s}$ have quantum dimension $\sqrt{2}$ and are ``descendants" of  $\pi$-flux vortices of the surface superconductor. The anyon $f e^{4 i \varphi_\rho}$ has electric charge $Q = 1$ and braids trivially with all the other anyons of the Moore-Read$\times U(1)_{-2}$ theory: it is identified with the physical electron.

Even though the intrinsic topological content of the surface state can be conveniently expressed as a direct product of a Moore-Read theory and a neutral anti-semion theory, the two sectors are linked by the time-reversal symmetry in a non-trivial manner. The action of ${\cal T}$ on the anyons is,
\bea {\cal T}:\quad &&\sigma e^{i m \varphi_\rho}  \to \sigma e^{i m \varphi_\rho} \bar{s}, \quad\quad \sigma e^{i m \varphi_\rho}  \bar{s} \to \sigma e^{i m \varphi_\rho}, \quad \quad \quad \quad m = 1, 3, 5, 7 \nn\\
&&e^{i m \varphi_\rho} \to e^{i m \varphi_\rho}, \quad\quad\quad {\cal T}^2 = +1, \quad\quad\quad \quad \quad\quad \quad\quad~\,m = 0, 4 \nn\\
&&f e^{i m \varphi_\rho} \to f e^{i m \varphi_\rho}, \quad \quad{\cal T}^2 = -1, \quad\quad\quad \quad \quad \quad \quad \quad \,\,m = 0, 4 \nn\\
&&e^{i m \varphi_\rho} \bar{s} \to e^{i m \varphi_\rho} \bar{s} , \quad \quad\,\, {\cal T}^2 = +1,\quad\quad\quad\quad\quad\quad\quad\quad~ m  = 2, 6 \nn\\
&&f e^{i m \varphi_\rho} \bar{s} \to f e^{i m \varphi_\rho} {\bar{s}}, \quad ~{\cal T}^2 =-1,  \quad\quad\quad\quad\quad\quad\quad\quad\,m  = 2, 6 \nn\\
&&e^{i m \varphi_\rho} \bar{s} \to f e^{i m \varphi_\rho}\bar{s}, \quad\quad  f e^{i m \varphi_\rho} \bar{s} \to  e^{i m \varphi_\rho}\bar{s},\quad\quad\quad \quad\,\, m  = 0, 4 \nn\\
&&e^{i m \varphi_\rho}  \to f e^{i m \varphi_\rho}, \quad\quad  ~~f e^{i m \varphi_\rho} \to  e^{i m \varphi_\rho},\quad\quad\quad \quad\quad~\,\, m  = 2, 6
\label{eq:TT24}\eea
For the anyons which are mapped to themselves under ${\cal T}$, we have noted the ``Kramers parity", ${\cal T}^2$.

The Moore-Read$\times U(1)_{-2}$  phase is separated from the superconducting surface phase by a continuous surface phase transition in the $XY^*$ universality class. A detailed understanding of this transition, including the fate of vortices across the critical point, makes us confident that
the Moore-Read$\times U(1)_{-2}$ phase can be realized on the ETI surface. As already noted, this phase transition is driven by the proliferation of flux $4 \pi$ vortices, when approached from the superconducting side. On the other hand, when approached from the topologically-ordered side, the phase transition is triggered by the condensation of the charge $1/2$ boson, $ e^{2 i \varphi_\rho} \bar{s}$, which corresponds to the ``elementary" field $\psi$ of the $XY$ model.  The star in $XY^*$ serves to remind that the local charge $2$ Cooper pair order parameter $cc$ is the fourth power of the $XY$ field, $cc \sim \psi^4$. 

We will demonstrate that a strictly 2d state with the Moore-Read$\times U(1)_{-2}$ topological content and electric charge assignments (\ref{Eq:MooreRead}) must have $\sigma_{xy} \neq 0$, and so is incompatible with time-reversal symmetry. However, as a surface phase of an ETI, the Moore-Read$\times U(1)_{-2}$ state is time-reversal invariant and carries $\sigma_{xy} = 0$; we will show that this is fully consistent with the bulk electromagnetic response of an ETI. Further, if we give up either the time-reversal symmetry or the $U(1)$ symmetry, the Moore-Read$\times U(1)_{-2}$ phase can be realized strictly in 2d (with the corresponding quantum numbers (\ref{Eq:MooreRead}) or (\ref{eq:TT24}) under the unbroken symmetry). This is obvious in the case when time-reversal is given up:  to obtain a 2d realization, stack together the well-known Moore-Read state of electrons with $\sigma_{xy} = 1/2$ and the $U(1)_{-2}$ state of neutral bosons. Let us label the resulting 2d phase of matter, ${\cal C}_{1/2}$. 

It is instructive to obtain the same 2d state, ${\cal C}_{1/2}$, in the following way. Imagine an ETI slab with a large but finite thickness. Place the top surface of the slab into the topologically-ordered symmetry-respecting phase and the bottom surface into the topologically trivial, ${\cal T}$-broken phase with $\sigma_{xy} = 1/2$. Since the bulk and the bottom surface of the slab have no intrinsic topological order, the whole slab viewed as a 2d system carries the topological order of the top surface and has a Hall-conductivity, $\sigma_{xy} = 1/2$. The identification of such a slab  with the 2d state ${\cal C}_{1/2}$ discussed above implies that the two systems have the same 1d edge states. This means that an edge on the ETI surface between the Moore-Read$\times U(1)_{-2}$ phase and the $\sigma_{xy} = -1/2$ phase is identical to the edge of ${\cal C}_{1/2}$ and supports a right-moving neutral Majorana mode $f$ (with central charge $c = 1/2$), a right-moving bosonic charge mode $e^{i \varphi_\rho}$ ($c = 1$), and a left-moving bosonic neutral mode $\bar{s}$ ($c = -1$). 

It may not be a priori obvious that a time-reversal invariant Moore-Read$\times U(1)_{-2}$ state, with ${\cal T}$ acting according to (\ref{eq:TT24}) (and $U(1)$ symmetry given up), can be realized in 2d. However, such a state can be constructed by using an ETI slab, whose top surface is in the topologically-ordered phase and the bottom surface in the superconducting phase. We will propose a possible route to explicitely realize an equivalent 2d state within an exactly solvable lattice model. This serves as an additional consistency check on our construction. We will also argue that the edge on the ETI surface between the topologically-ordered phase and the superconducting phase is generally gapped. 

The fact that electrons transform as Kramers doublets under ${\cal T}$ plays an important role throughout our investigation of the surface properties of an ETI. This makes us wonder if there are any
non-trivial electron insulators in 3d with ${\cal T}^2 = +1$. We know that there are no such phases within the non-interacting realm.\cite{Ludwig,KitaevFree} It is, however, not immediately clear whether
there is any true obstruction for such phases to exist once strong interactions are present. More specifically, as we have already noted, ordinary ETIs with ${\cal T}^2 = (-1)^N$ can be distinguished from
trivial electron insulators by their non-zero $\theta$-parameter. We remind the reader that the 
$\theta$-variable is periodic modulo $2\pi$ and transforms as $\theta \to -\theta$ under ${\cal T}$. Thus, the distinct, time-reversal invariant values of $\theta$ are $\theta = 0$ and $\theta = \pi$.
Trivial electron insulators have $\theta = 0$, while standard ETIs with ${\cal T}^2 = (-1)^N$ have $\theta = \pi$. At a ``classical level," there is no connection between $\theta = \pi$ and 
${\cal T}^2 = (-1)^N$, and so one might ask if (interacting) time-reversal invariant electron insulators with $\theta = \pi$ and ${\cal T}^2 = +1$ exist. In this paper, we will show that the answer to this question is negative: at the quantum level, $\theta = \pi$ is compatible with time-reversal invariance only if ${\cal T}^2 = (-1)^N$. Of course, we cannot rule out the existence of non-trivial interacting electron insulators with $\theta = 0$ and ${\cal T}^2 = +1$.

This paper is organized as follows. In section \ref{sec:superfluid}, we discuss the properties of the superconducting surface phase of an ETI. We pay particular attention to the statistics of flux-tubes on the superconducting surface. Section \ref{sec:cond8pi} discusses the gapped symmetry-preserving topologically-ordered phase obtained from the superconductor via the condensation of flux $8\pi$ vortices. This phase has $96$ anyon types; we label it, $T_{96}$. For pedagagoical reasons, we find it simpler to discuss this ``larger" phase, before turning to the Moore-Read$\times U(1)_{-2}$ phase obtained by condensing flux $4 \pi$ vortices. Section \ref{sec:Timp} discusses the implementation of time-reversal symmetry in the $T_{96}$ phase. Section \ref{sec:phicond} discusses the phase transition between the $T_{96}$ phase and the superconducting surface phase in detail. Section \ref{sec:reduced} obtains the ``smaller" Moore-Read$\times U(1)_{-2}$ surface phase from the $T_{96}$ phase by condensing a bosonic anyon, which is a ``descendant" of a flux $4\pi$ superconducting vortex. Section \ref{sec:Witten} demonstrates the consistency of the topologically-ordered surface phases $T_{96}$ and Moore-Read$\times U(1)_{-2}$ with the Witten effect in the ETI bulk. Section \ref{sec:topoSC} is devoted to the physics of an ETI slab with the topologically-ordered phase on the top surface and the superconducting phase on the bottom surface. Section \ref{sec:SCstar} discusses a superconducting phase, SC$^*$, obtained from the Moore-Read$\times U(1)_{-2}$ surface phase by condensing a charge $1$ boson, $e^{4 i \varphi_\rho}$. This phase carries a remnant Abelian topological order (thus the superscript). Various loose ends are delegated to the appendix. In particular, we draw the reader's attention to appendix \ref{app:thetapi}, which demonstrates the incompatibiltiy of ${\cal T}^2 = +1$ and electromagnetic response with $\theta = \pi$.

\section{Superconducting surface.}
\label{sec:superfluid}
Our strategy will be to start with the superconducting surface of an electron topological insulator, which preserves
time-reversal symmetry but breaks  particle number conservation. We will then restore the particle number
symmetry by ``condensing" superconducting vortices, obtaining a fully gapped time-reversal invariant
topologically-ordered surface state.  To implement this strategy, we found that a useful technical innovation is to consider a modified theory in which the $U(1)$ symmetry is gauged.    In the gauged theory the Abelian statistical phases associated with vortices are well defined.  While we emphasize that our ultimate object is to describe the ungauged vortex-condensed topologically-ordered state, the gauged theory offers a convenient method for deducing the statistical phases of the anyons of the topologically ordered phase that are derived from vortices.

As a first step we need to understand the properties of the superconducting surface. We infer these 
by coupling the single Dirac fermion on the surface to the superconducting order parameter and solving the resulting
Bogoliubov-de-Gennes (BdG) equation.\cite{FuKaneMajorana} In the absence of vortices in the order parameter, the surface superconductor has the following
excitations: i) a gapped fermionic Bogolioubov quasiparticle $f_{\sigma}$, which transforms as a Kramers doublet under time-reversal symmetry;
ii) a gapless Goldstone mode, which will play little role in our discussion below. 

The superconducting vortices have the following properties: a vortex with odd
vorticity carries a single Majorana zero mode, while a vortex with even vorticity carries no zero mode. In our notation, a vortex with vorticity $k$ carries
magnetic flux $\Phi = \pi k$ once the $U(1)$ particle number symmetry is gauged. (The presence/absence of zero-modes is independent of whether the $U(1)$
symmetry is gauged.) Thus, the ground state of a vortex with even
vorticity is unique, as are the excited states, which can be obtained by adding Bogolioubov quasiparticles. Note that the state obtained by adding an odd number of Bogoliubov quasiparticles
 to such an ``even" vortex is not a Kramers doublet, since the vorticity breaks the time-reversal symmetry: the time-reversal partner has the opposite vorticity.

As for vortices with odd vorticity, the presence of the Majorana zero mode implies that two such vortices separated by a large distance have a 2-fold degenerate ground state, obtained
by either leaving empty or filling the complex fermion zero mode formed out of the two Majorana zero modes. The physics is analogous to that in a $p_x+ip_y$ superconductor (however, we will
discuss the important distinction shortly).  Thus, vortices with odd vorticity have non-Abelian  statistics. If the $U(1)$ symmetry is not gauged, this statistics is only ``projective," since
the logarithmic interactions between vortices make the Abelian part of the exchange statistics ill-defined. On the other hand, if the $U(1)$ symmetry is gauged, the Abelian part
of the exchange statistics becomes meaningful. We now gauge the $U(1)$ symmetry with a weakly fluctuating electromagnetic gauge field $A_{\mu}$ to expose this Abelian part of the statistics. This 
procedure is only a technical trick - ultimately, we are interested in the physics in the absence of the fluctuating electromagnetic field.

A classical solution to the Maxwell's equations for a static flux $\Phi$ flux-tube on the $z = 0$ surface of a semi-infinite topological insulator has
the form,
\beq \vec{B} = \frac{\Phi}{2 \pi} \frac{(x, y, |z|)}{r^3} \label{eq:Bmon}\eeq
Thus, the flux is spread out in the bulk of the insulator (and in the vacuum outside), but localized on the surface. This spreading out of the flux leads to a $1/r$ ``diamagnetic" interaction between flux-tubes.
This interaction decays quickly enough to make the flux-tube statistics well-defined.\footnote{Strictly speaking, the statistical interaction and the diamagnetic interaction decay at the same rate.}

To determine the Abelian part of the flux tube statistics we use the following argument. First, let us recall that another surface termination of the electron topological insulator is provided
by a fully gapped state with no intrinsic topological order, which preserves the particle-number symmetry, but spontaneously (or explicitly) breaks the time-reversal symmetry and carries a Hall-conductivity $\sigma_{xy} = \pm 1/2$. 
Such a state can be obtained by applying a Zeeman field on the surface, which induces  a mass term for the surface Dirac fermion. As can be easily checked by solving the BdG equation on the surface, the edge between the superconducting phase and the $\sigma_{xy} = -1/2$ phase carries a single gapless chiral Majorana mode (central charge $c =1/2$). The chirality is right-moving (counter-clockwise) for a superconducting droplet in a $\sigma_{xy}= -1/2$ phase.

Now, imagine a slab of the topological insulator with thickness $d$ in the $z$ direction much greater than the lattice spacing $a$. We further take the length and width
of the slab to be much greater than $d$. We take the top surface of the slab to be in the superconducting phase and the bottom surface in the $\sigma_{xy} = 1/2$ phase, with the sign of $\sigma_{xy}$ defined, as usual,
relative to the $\hat{z}$ axis (i.e. the Hall-conductivity is $\sigma_H = -1/2$ with respect to the outward normal $-\hat{z}$ of the insulator surface). The edge of the slab will carry a single right-moving Majorana mode. 
The slab viewed as a 2d system is identical to a $p_x+ip_y$ superconductor. Indeed,  let us start with a non-interacting electron topological insulator slab and explicitly break the particle-number symmetry on the top surface and time-reversal on the bottom surface. The slab is a 2d non-interacting superconductor with no global symmetries (except the particle-hole symmetry of the BdG equation) - i.e. it belongs to class {\bf D}. According to the classification of Refs.~\onlinecite{Ludwig,KitaevFree}, 2d phases in class {\bf D} are labeled by an integer $n$ and are equivalent to $n$ copies of a $p_x+ip_y$ superconductor. A system in phase $n$ supports $n$ right-moving Majorana modes on the edge. Therefore, our slab with a single Majorana mode on the edge, viewed as a 2d system, is a $p_x+ip_y$ superconductor.

Now, let us gauge the $U(1)$ particle-number symmetry and consider a flux-tube piercing our 2d slab. As already mentioned, for a magnetic field configuration satisfying the classical Maxwell equations, the magnetic flux on the superconducting surface is concentrated in the vortex core, but is spread out in the bulk of the slab and on the bottom (insulating) surface. The exact details of the flux distribution, except its smoothness in the bulk and on the bottom surface, will play no role in our discussion below; for instance, instead of using the solution to Maxwell's equations, we can just impose a flux distribution which is uniform along the $z$ direction and has a characteristic radius $R \gg a$. We can now ask about the statistics of the flux-tubes piercing the slab. If the distance between the flux-tubes is much larger than their radius, we can view the system as two-dimensional. The statistics then become identical to statistics of flux-tubes in a gauged $p_x+ip_y$ superconductor. These were discussed in Refs.~\onlinecite{ReadGreen, Ivanov2001, Stern04, Stone06, KitaevHoneycomb, Read2008} and are identical to those in the Ising anyon model, which we briefly review below.

Recall that the Ising anyon model has the following anyon types: $1, \sigma, f$. The fusion rules are $f \times f = 1$, $\sigma \times \sigma = 1 + f$ and $\sigma \times f = \sigma$. The topological spins $\theta$ and the quantum dimensions $d$ are listed in Table \ref{tbl:Ising}. The topological spins determine the mutual (full braid) statistics of quasiparticles $a$ and $b$ fused in channel $c$ to be $M^{a b}_c = \frac{\theta_c}{\theta_a \theta_b}$. (Here and below we always give statistics for a counter-clockwise exchange). The mutual statistics can also be expressed through the $R$ symbols as $M^{a b}_c = R^{a b}_c R^{b a}_c$, where $R^{a b}_c$ denotes the phase picked up during an exchange of anyons $a$ and $b$ fused in channel $c$. For $a \neq b$, $R^{a b}_c$ does not have a gauge-invariant meaning (while $M^{a b}_c$ does), but for $a = b$, $R^{a a}_c$ gives the self-statistics of anyons $a$ (fused in channel $c$).  We  list these self-statistics $R^{aa}_c$ in Table \ref{tbl:Ising}.

\begin{table}
\begin{tabular}{|c|c|c|c|}
\hline
\,\, & $1$ & $\sigma$ & $f$\\
\hline
$d$ & $1$ & $\sqrt{2}$ & $1$\\
\hline
$\theta$ & $1$ & $e^{\pi i /8}$ & $-1$ \\
\hline
$R$ & $R^{1 1}_{1} = 1$ & $R^{\sigma \sigma}_1 = e^{-\pi i /8}$ & $R^{ff}_1  = -1$\\
& & $R^{\sigma \sigma}_f = e^{3 \pi i /8}$ & \\
\hline
\end{tabular}
\caption{Quantum dimensions $d$, topological spins $\theta$  and self-statistics $R$ of anyons in the Ising model.}
\label{tbl:Ising}
\end{table}

Returning to the excitations of a $p_x+ip_y$ superconductor, fermionic Bogolioubov quasiparticles naturally belong to the $f$ sector. Flux-tubes with odd vorticity belong to the $\sigma$ sector, while flux-tubes with even
vorticity belong to the $1$ sector or the $f$ sector (as already noted, for an even flux tube, one can toggle between these sectors by exciting a Bogolioubov quasi-particle). Note that given an even flux-tube (e.g. with flux $2\pi$) in a 2d system, the flux can continuosly shrink to pass through a single plaquette, becoming invisible. Thus, an even flux-tube has to lie in the same sector as excitations with no flux, i.e. precisely $1$ or $f$. 

Thus, we know the statistics of flux-tubes passing through the entire 2d slab. Clearly, these statistics are sensitive to the $\sigma_{xy} =1/2$ phase that we placed on the bottom surface of the slab. In particular, they
explicitely break the time-reversal symmetry. If we instead placed the $\sigma_{xy} = -1/2$ phase on the bottom surface, the slab as a whole would behave as a $p_x-ip_y$ superconductor and the statistics would be time-reversal conjugates (i.e. complex conjugates). We would now like to separate out the contribution to the flux-tube statistics coming from the top (superconducting) surface. This contribution should be i) time-reversal invariant, ii) independent of what phase the bottom surface is in.

Consider the effective action for the slab as a whole, $S_{\mathrm{slab}}$, describing the motion of flux tubes and Bogolioubov quasiparticles. On one hand, as already noted,
this motion is governed by the Ising anyon theory $S_{\mathrm{slab}} = S_{\mathrm{Ising}}$.  On the other hand, we may decompose $S_{\mathrm{slab}}$ as,
\beq S_{\mathrm{slab}} = S_{\mathrm{top}}[j_v, j_f] + S_{\mathrm{bulk}}[A] + S_{\mathrm{bottom}}[A] \label{eq:Stbb} \eeq
Here, $S_\mathrm{top}$, $S_{\mathrm{bulk}}$ and $S_{\mathrm{bottom}}$ are the actions for the top surface, bulk and bottom surface, respectively. $j_v$ and $j_f$ are the vortex and Bogolioubov quasiparticle currents on the top surface. The bulk and the bottom surface are affected during the vortex motion only through the electromagnetic gauge field $A_{\mu}$, which we take to be a classical, adiabatically varying background field, slaved to the vortex coordinates. Thus, to compute $S_{\mathrm{bulk}}[A_{\mu}]$ and $S_{\mathrm{bottom}}[A_{\mu}]$ we may integrate the electrons out. The bulk response gives the usual Maxwell term
\beq S_{\mathrm{bulk}} \sim \int d^3x d \tau F^2_{\mu \nu}\eeq
This term contributes to the aforementioned $1/r$ diamagnetic interactions between the flux tubes and is irrelevant for our purposes. For the bottom surface we obtain an effective Chern-Simons (CS) action,
\beq S_{\mathrm{bottom}} = -\frac{i k}{4\pi} \int_{\mathrm{bott}} d^2 x d \tau \epsilon_{\mu \nu \lambda}  A_{\mu} \d_{\nu} A_{\lambda}, \quad\quad k = 1/2 \label{eq:Sbott} \eeq
at level $k = \sigma_{xy} = 1/2$. We note that as far as the bulk electromagnetic response of the electron topological insulator is concerned, it is often stated that a $\theta$ term,
\beq S_\theta = -\frac{i \theta}{32 \pi^2} \int_{\mathrm{bulk}} d^3x d\tau \epsilon_{\mu \nu \lambda \sigma} F_{\mu \nu} F_{\lambda \sigma}\label{eq:Stheta}\eeq
with $\theta = \pi$ is present. However, in the absence of monopoles in the bulk, this $\theta$ term reduces to a CS term on the boundary,
\beq S_\theta = -\frac{i \theta}{8 \pi^2} \int_{\mathrm{bound}} d S_{\mu}  \epsilon_{\mu \nu \lambda \sigma} A_{\nu} \d_{\lambda} A_{\sigma} \label{eq:Sbyparts}\eeq
In Eq.~(\ref{eq:Stbb}) we choose to incorporate this boundary term into the actions for the top and bottom surfaces $S_{\mathrm{top}}$ and $S_{\mathrm{bottom}}$. Indeed, 
for the choice $\theta = \pi$, the action for the bottom surface in Eq.~(\ref{eq:Sbott}) is precisely given by the boundary contribution in Eq.~(\ref{eq:Sbyparts}). On the other hand, the action 
for the top surface $S_{\mathrm{top}}$ is more complicated (in fact, due to the presence of non-perturbative vortex configurations carrying Majorana zero modes, 
the top surface cannot be described by a simple quadratic theory for the gauge field $A_{\mu}$), however, whatever the form of this action, the bulk $\theta$-term is accounted for
in it.

From Eq.~(\ref{eq:Stbb}) we isolate the action for the top surface alone:
\beq S_{\mathrm{top}}[j_v, j_f] = S_{\mathrm{Ising}}[j_v, j_f] - S_{\mathrm{bottom}}[A] \label{eq:Stop1}\eeq
(Here we've dropped the non-topological bulk Maxwell term). It is convenient to rewrite the action for the bottom surface in terms of the vortex currents. The gauge field $A_{\mu}$ is slaved to the vortex currents and for well-separated vortices, we may write $\epsilon_{\mu \nu \lambda} \d_{\nu} A^{\mathrm{bottom}}_{\lambda} = \pi j_{v,\mu}$.
Enforcing this constraint with a Lagrange multiplier $a_{\mu}$ we obtain,
\beq S_{\mathrm{top}} =S_{\mathrm{Ising}}[j_v, j_f]+ \int d^2 x d\tau \left[\frac{i}{8\pi} \epsilon_{\mu \nu \lambda} A_{\mu} \d_{\nu} A_{\lambda} + i a_{\mu} \left(\frac{1}{\pi} \epsilon_{\mu \nu \lambda} \d_{\nu} A_{\lambda} - j^v_{\mu}\right)\right]\eeq
Integrating over the (now unconstrained) gauge field $A_{\mu}$, 
we obtain the desired action for the top surface,
\beq  S_{\mathrm{top}} = S_{\mathrm{Ising}}[j_v, j_f]  + \int d^2 x d\tau \left(\frac{-8 i}{4\pi} \epsilon_{\mu \nu \lambda} a_{\mu} \d_{\nu} a_{\lambda} - i a_{\mu} j^v_{\mu}\right)\label{eq:Stop2}\eeq
The second term in Eq.~(\ref{eq:Stop2}) is a CS theory for the gauge field $a_{\mu}$ at level $k = -8$, $U(1)_{-8}$. The vortex carries charge $1$ under the CS field $a_{\mu}$.
The CS term contributes an extra phase to the Abelian statistics of vortices, in addition to the 
statistics coming from the Ising action in the first term. We can, thus, think of our vortices as embedded in an anyon model which is a direct product, $\mathrm{Ising}\times U(1)_{-8}$. 

We remind the reader that a general $U(1)_{k}$ anyon model (with $k$ - even) has $|k|$ distinct anyon types, which we denote here by $e^{i l \theta}$, $l = 0 \ldots |k|-1$. One may think of $l$ as an integer modulo $|k|$. The anyons have Abelian fusion rules:
$e^{i l_1 \theta} \times e^{i l_2 \theta} =  e^{i (l_1 + l_2) \theta}$ and topological spins $\theta_l = e^{\pi i l^2/k}$. As with all Abelian anyons, the topological spins are equal to the self-statistics, $R^{ll}_{2l} = \theta_l$. 

Now, from Eq.~(\ref{eq:Stop2}), we associate  flux tubes on the surface of an ETI with anyons of the $\mathrm{Ising}\times U(1)_{-8}$ theory as follows. Labeling the vorticity as $k$,  flux tubes with $k$ - odd, correspond to $\sigma e^{i k \theta}$ anyons, while flux-tubes with $k$-even correspond to $e^{i k \theta}$ and $f e^{i k \theta}$ anyons. Here, ${1, \sigma, f}$ labels run over the Ising part and $e^{i k \theta}$ - over the $U(1)_{-8}$ part.  The Bogolioubov quasiparticle (electron) lies in the zero flux sector and is just $f$. As noted before, the $U(1)_{-8}$ charge coincides with the vorticity. Note that the vorticity (flux) on the surface of an ETI is conserved unless a monopole of $A_{\mu}$ passes through the surface. For now, we do not consider such monopole events, thus, for the present purposes it is appropriate to think of the vorticity label $k$ as an integer, rather than an integer modulo $8$. The statistical properties, however, are periodic under $k \to k +8$ (i.e. flux $\Phi \to \Phi + 8 \pi$). This is different from a strictly 2d system, where as we mentioned, the statistical properties of flux-tubes must be invariant under $\Phi \to \Phi + 2 \pi$. 

We note that not all anyon types of the $\mathrm{Ising}\times U(1)_{-8}$ theory are realized by the flux tubes on the surface. Namely the $e^{i k \theta}$, $f e^{i k \theta}$ anyons with $k$-odd and $\sigma e^{i k \theta}$ anyons with $k$ - even are absent. We will refer to the allowed anyon types together with their fusion and braiding rules as the ``restricted" $\mathrm{Ising}\times U(1)_{-8}$ theory. We note that the allowed anyon types are analogous to those of the Moore-Read (Pffafian) state, whose quasiparticles form a subset of the $\mathrm{Ising} \times U(1)_{+8}$ theory. 

The topological spins $\theta$ of the allowed flux-tubes are listed in Table \ref{tbl:vorts}. Here we use the fact that the topological spin $\theta$ of an anyon in the product theory is the product of topological spins of the constituents. The same holds for the $R$ matrix elements. The $R$ matrix elements describing the self-statistics of $\sigma e^{i k \theta}$ flux-tubes are, thus, $R^{\sigma_k \sigma_k}_{1_{2k}} = e^{- \pi i (k^2 + 1)/8}$ and
$R^{\sigma_k \sigma_k}_{f_{2k}} = e^{- \pi i (k^2 - 3)/8}$. (Here and below, we will use the short-hand notation $\alpha_k = \alpha e^{i k \theta}$). Since all the other flux-tubes are Abelian, their self-statistics is just given by the topological spin.

\begin{table}
\begin{tabular}{|c|c|c|c|c|c|c|c|c|}
\hline
\,\, &$0$ & $1$ & $2$ & $3$ & $4$ & $5$ & $6$ &$7$\\
\hline
$1$ & $1$ & \,\, & $-i$ & \,\, & $1$ & \, \, & $-i$ & \, \, \\
\hline
$\sigma$ & \,\, & $\,1\,$ & \,\, & $-1$ & \, \, & $-1$ & \, \,& $\,1\,$ \\
\hline
$f$ &$-1$ & \,\, & $i$ & \,\, & $-1$ & \, \, & $i$ & \, \, \\
\hline
\end{tabular}
\caption{Topological spins $\theta$ of  flux-tubes on the surface of an ETI. The statistics of flux-tubes are identical to those of anyons in an 
$\mathrm{Ising}\times U(1)_{-8}$ theory. The $\mathrm{Ising}$ content is specified in the row label and the $U(1)_{-8}$ content, which is identical 
to the vorticity $k$, in the column label. The blank anyons are not allowed. The statistics are invariant under $k \to k + 8$. }

\label{tbl:vorts}
\end{table}

Let us now discuss the transformations of our flux-tubes under time-reversal symmetry. Since we have inferred the statistical properties by starting with
a 2d slab, whose lower surface explicitely broke ${\cal T}$, time-reversal symmetry is not manifest in the labeling that we are using for the flux-tubes.
Nevertheless, we can work out the action of time-reversal symmetry by noting that i) ${\cal T}$ maps flux-tubes with vorticity $k$ to flux-tubes with vorticity
$-k$, ii) topological spins of time-reversal partners must be complex-conjugates. These requirements uniquely fix the action of time-reversal symmetry to be:

\bea {\cal T}:\quad &&\sigma e^{i k \theta} \to \sigma e^{-i k \theta}, \,\,\,\,\quad\quad\quad \quad\quad\quad\quad k \,-\, \mathrm{odd} \nn\\
&& e^{i k \theta} \to e^{-i k \theta}, \quad f e^{i k \theta} \to f e^{-i k \theta}, \quad k \equiv 0\,\, (\mathrm{mod} \,\,4) \nn\\
&& e^{i k \theta} \to fe^{-i k \theta}, \quad f e^{i k \theta} \to e^{-i k \theta}, \quad k \equiv 2\,\, (\mathrm{mod} \,\,4) \label{eq:Tvort}\eea
Furthermore, the trivial zero flux sector $1$ is a Kramers singlet, while the Bogolioubov quasiparticle $f$ is a Kramers doublet. As already noted,
all the other flux-tubes carry no additional degeneracy. One can check that the fusion rules and braidings are consistent with the time-reversal symmetry
(\ref{eq:Tvort}). To gain some intuition about the transformation properties (\ref{eq:Tvort}), we discuss several explicit examples.

First, let us consider the flux $2 \pi$ vortex $e^{2 i \theta}$. It has topological spin (self-statistics) $\theta_{1_2} = -i$. According to Eq.~(\ref{eq:Tvort}), the 
time-reversal partner of this vortex is a flux $-2 \pi$ vortex $f e^{-2 i \theta}$, whose topological spin is $\theta_{f_{-2}} = i$, which is the complex conjugate of $\theta_{1_2}$, 
as required. Note that the time-reversal partners $e^{2 i \theta}$ and $f e^{-2 i \theta}$ are mutual semions and fuse to the fermion $f$, which is a Kramers doublet. 
We show in Appendix \ref{app:Kramers} that two mutual semions, which are mapped into each other by time-reversal symmetry, actually, {\it must} fuse to a {\it Kramers doublet}  fermion.
The fact that the fusion product must be a fermion is easy to see as the mutual statistics of two anyons $a$ and $b$ fused in channel $c$ is $M^{ab}_c = \theta_c/(\theta_a \theta_b)$.
If $a$ and $b$ are time-reversal partners, $\theta_a\theta_b = 1$, so mutual semionic statistics implies $\theta_c = -1$. Note that the result of Appendix \ref{app:Kramers} also works when 
the time-reversal partners $a$ and $b$ are non-Abelian, assuming that these have semionic statistics in some specific Abelian channel $c$ and that $c$ transforms into itself under ${\cal T}$.

There actually exists a simple explicit example of a 2d topologically-ordered state where two mutual semions transform into each other under time-reversal symmetry and fuse to a fermion, which is
a Kramers doublet. This state is just a toric code where the mutual semions $e$ and $m$ are time-reversal partners and fuse to a Kramers doublet fermion $f$.
We will discuss this toric code in more detail in section \ref{sec:SCstar}.

Now, let us turn our attention to flux $\pm\pi$ vortices $\sigma e^{i \theta}$ and $\sigma e^{-i \theta}$. These  time-reversal partners both have topological spin $\theta_{\sigma_{\pm 1}} = 1$ and
fuse according to $\sigma e^{i \theta} \times \sigma e^{- i \theta} = 1 + f$. When fused in channel $1$, the two flux-tubes have bosonic mutual statistics and, thus, give rise to a Kramers singlet $1$. On the other
hand, when fused in channel $f$, they have semionic mutual statistics and give rise to the Kramers doublet $f$, in accordance with the result of Appendix \ref{app:Kramers}. Now, let us consider the self-statistics
of $\sigma e^{i \theta}$. This flux-tube fuses with itself to give $\sigma e^{i \theta} \times \sigma e^{i \theta} = e^{2 i \theta} + f e^{2 i \theta}$.  On the other hand, its time-reversal partner $\sigma e^{-i\theta}$ fuses with itself as $\sigma e^{- i \theta} \times \sigma e^{-i \theta} = e^{-2 i\theta} + f e^{-2 i \theta}$. Now recall that the fusion products transform under ${\cal T}$ as $e^{2 i \theta} \leftrightarrow f e^{-2 i \theta}$ and $f e^{2 i \theta} \leftrightarrow e^{-2 i \theta}$. The self-statistics $R^{\sigma_1 \sigma_1}_{1_2} = \left(R^{\sigma_{-1} \sigma_{-1}}_{f_{-2}}\right)^* = e^{-\pi i /4}$ and $R^{\sigma_1 \sigma_1}_{f_2} = \left(R^{\sigma_{-1} \sigma_{-1}}_{1_{-2}}\right)^* = e^{\pi i /4}$ are consistent with this. 

Before we conclude this section, we would like to stress that the restricted $\mathrm{Ising} \times U(1)_{-8}$ theory that we discussed in this section describes the statistical properties of flux-tubes on the superconducting surface of an ETI. It {\it does not} describe the symmetry-respecting topologically-ordered surface of an ETI, which is the subject of the next section. In the absence of a dynamical 
external electromagnetic field, the superconducting surface has no intrinsic topological order and its vortices are not local excitations. The introduction of a weakly fluctuating electromagnetic gauge field is just a useful technical trick on route to exposing the properties of the symmetry-respecting surface with intrinsic topological order. We will see that one version of this topological order contains all the anyons of the restricted $\mathrm{Ising}\times U(1)_{-8}$ theory, as well as some additional anyons.

One can ask an independent question of whether a strictly 2d fermion system can support the restricted $\mathrm{Ising} \times U(1)_{-8}$ topological order with the above implementation of ${\cal T}$. The same question was raised with regards to the same intrinsic topological order, but with a slightly different implementation of ${\cal T}$ in Refs.~\onlinecite{BNQ,AshvinETI}.  In both cases, the question is currently under investigation. 

\section{Symmetry-respecting surface topological order.}
\subsection{Vortex Condensation.}
\label{sec:cond8pi}
We are now ready to quantum disorder the surface superconductor by proliferating vortex defects. As a result, we will obtain an insulating time-reversal-respecting surface with intrinsic topological order.

As a first step, we switch off the fluctuating electromagnetic field $A_{\mu}$ that was introduced in the previous section. We, nevertheless, continue to label the surface vortices by their anyonic type (statistics) in the gauged theory. We use the following prescription for condensing vortices. We will justify this prescription in section \ref{sec:phicond}. Only Abelian vortices (those with quantum dimension $d = 1$) with bosonic self-statistics  (as flux-tubes) can condense. The minimal Abelian bosonic vortex has flux $4\pi$ - it is $e^{4 i \theta}$ in the notation of the previous section. Although we understand how to condense this vortex, we find it conceptually simpler to begin by condensing a flux $8 \pi$ vortex $e^{8 i \theta}$. The reason for this is that $e^{8 i \theta}$ has trivial mutual statistics with all the other excitations. On the other hand, $e^{4 i \theta}$ has semionic mutual statistics with odd vortices. We will come back to discuss the condensation of $e^{4i\theta}$ in section \ref{sec:reduced}, but for now, focus on condensing $e^{8 i \theta}$.

We claim that the proliferation of $e^{8 i \theta}$ vortices gives rise to a topologically-ordered state with the following properties. 
\vspace{0.2cm}

i) The particle number conservation symmetry is restored.

ii) An electric charge $1/4$ Abelian bosonic anyon $e^{i \phi}$ appears in the spectrum. This charge ``quantum" is dual to the flux $8\pi$ of the condensing vortex. 

iii) Vortices with vorticity $0 \le k < 8$ emerge out of the superconductor as {\it electrically neutral} anyons, with the same fusion and braiding rules, which they possessed as flux-tubes. We continue to label the resulting anyons as $\sigma e^{i k \theta}$, $k = 1, 3, 5, 7$, and $e^{i k \theta}$, $f e^{i k \theta}$, $k = 0, 2, 4, 6$. As the flux $8 \pi$ vortex, $e^{8 i \theta}$, is condensed, $k$ is now truly an integer modulo $8$. Note that the zero vorticity sector gives rise to the vacuum anyon $1$ and a neutral fermion $f$, which is the descendant of the Bogolioubov quasiparticle.

iv) The charge $1/4$  boson $e^{i \phi}$ has mutual statistics $e^{-i \pi k /4}$ with the $\{1, \sigma, f\} e^{i k\theta}$ anyons.
\vspace{0.2cm}

All the anyons of the topologically-ordered state can be obtained by fusing the descendants of vortices with some number $m$ of charge $1/4$ bosons, $e^{i m \phi}$.  Thus, the anyon content of the surface topologically-ordered state is given by
\bea &&\sigma e^{i m \phi} e^{i k \theta}, \quad\quad \quad \quad \quad \quad \,\,\,\,\,\, 0 \leq m \leq 7,\,\, k = 1, 3, 5, 7,\,\, \quad\quad Q = m/4 \nn\\
&& e^{i m \phi} e^{i k \theta},\,\, fe^{i m \phi} e^{i k \theta}, \quad \quad\quad 0 \leq m \leq 7,\,\, k = 0, 2, 4, 6,\,\, \quad \quad Q = m/4
\label{Eq:TopoLarge} \eea
Here we've also listed the electric charge $Q$ of the anyons. We note that $e^{8 i \phi}$ is a charge $2$ boson, which braids trivially with all the other anyons. We identify this boson with the electron Cooper pair. Since the Cooper pair is a local bosonic microscopic excitation, we do not
count $e^{8 i \phi}$ as a separate anyon. 
 We further observe that $f e^{4 i \phi}$ is a charge $1$ fermion, which braids trivially with all the other anyons. It is the only anyon with this property,  we, therefore, identify it with the physical electron $c$.  
The surface topological order obtained has $96$ anyons, counting the electron, and $48$ anyons if we identify anyons which differ by an electron. The latter count is relevant for computing the ground state degeneracy
of the system on a (solid) torus. We label the present topological order as $T_{96}$.  In section \ref{sec:reduced} we will discuss a phase transition which reduces the number of anyons by a factor of $4$. 

We can write the following ``schematic" Lagrangian for the topologically-ordered surface,

\bea L_{\mathrm{topo}} &=& L_{\mathrm{Ising}}[j_\theta, j_f]  +  \frac{-8 i}{4\pi} \epsilon_{\mu \nu \lambda} a_{\mu} \d_{\nu} a_{\lambda} - i a_{\mu} j^{\theta}_{\mu} \nn\\
&+& \frac{i}{4 \pi} \epsilon_{\mu \nu \lambda}  \alpha^T_{\mu} K \d_{\nu} \alpha_\lambda - i \alpha^T_{\mu} J_{\mu} - \frac{i}{4} A_{\mu} j^{\phi}_{\mu} \label{eq:Ltopo1}\eea
with $\alpha_{\mu} = (\alpha^1_{\mu}, \alpha^2_{\mu})$, $J_{\mu} = (j^{\phi}_{\mu}, j^{\theta}_{\mu})$ and $K = -\left(\begin{array}{cc} 0 & 8\\8 & 0\end{array}\right)$.
The first line in Eq.~(\ref{eq:Ltopo1}) is identical to the effective action for flux-tubes in the superconducting phase, Eq.~(\ref{eq:Stop2}), with the renaming of the vortex current $j_v$ into $j_\theta$.
The second line encodes the mutual $e^{-\pi i/4}$ statistics between the $e^{i \phi}$ anyons, whose current is denoted by $j_\phi$, and $e^{i \theta}$. This mutual statistics is 
represented with the aid of a two-component CS gauge-field $\alpha_{\mu}$. We also include the coupling of the electromagnetic
gauge field $A_{\mu}$ to the charge $1/4$ anyons $e^{i \phi}$. It is easy to see that the ``Abelian part" of the statistics, which is encoded in Eq.~(\ref{eq:Ltopo1}) with the aid of three CS gauge fields,
$a$, $\alpha^{1}$, $\alpha^2$, can equally well be represented with just two CS gauge-fields, $\beta = (\beta^1, \beta^2)$ as,
\beq L_{\mathrm{topo}} = L_{\mathrm{Ising}}[j_\theta, j_f]  +\frac{i}{4 \pi} \epsilon_{\mu \nu \lambda}  \beta^T_{\mu} K' \d_{\nu} \beta_\lambda - i \beta^T_{\mu} J_{\mu} - \frac{i}{4} A_{\mu} j^{\phi}_{\mu} 
\label{eq:Ltopo2}\eeq
with $K' = \left(\begin{array}{cc} 8 & -8\\-8 & 0\end{array}\right)$. We next define $e^{i \varphi_\rho} = e^{i \phi} e^{-i \theta}$ and rewrite all the anyons in terms of $e^{i \varphi_\rho}$ and $e^{i \theta}$:
\bea &&\sigma e^{i m \varphi_\rho} e^{i k \theta}, \quad\quad \quad \quad \quad \quad \,\,\,\,\,\,\,\,\,\, k + m - \mathrm{odd},\,\, 0 \leq m \leq 7,\,\,0 \leq k \leq  7,\,\, \quad\quad Q = m/4 \nn\\
&& e^{i m \varphi_\rho} e^{i k \theta},\,\, fe^{i m \varphi_\rho} e^{i k \theta}, \quad \quad\quad k + m - \mathrm{even},\,\, 0 \leq m \leq 7,\,\, 0 \leq k \leq 7, \quad \quad Q = m/4 \nn\\
\label{Eq:TopoLargerho} \eea
Below, will use the two anyon labelings (\ref{Eq:TopoLarge}) and (\ref{Eq:TopoLargerho}) interchangably. We note that $e^{i \varphi_\rho}$ carries electric charge $1/4$, thus the subscript. In the language of the CS theory (\ref{eq:Ltopo2}) the relabelling generates an $SL(2,\mathbb{Z})$ transformation,
\beq \left(\begin{array}{c} j_\phi\\ j_\theta\end{array}\right) = \left(\begin{array}{cc} 1& 0\\-1 & 1\end{array}\right) \left(\begin{array}{c} j_\rho\\ \tilde{j}_\theta\end{array}\right) \eeq
where $j_\rho$, $\tilde{j}_\theta$ are the currents corresponding to  $e^{i \varphi_\rho}$ and $e^{i \theta}$ in the new ``basis." Eq.~(\ref{eq:Ltopo2}) then becomes,
\beq L_{\mathrm{topo}} = L_{\mathrm{Ising}}[\tilde{j}_\theta - j_\rho, j_f]  +\frac{i}{4 \pi} \epsilon_{\mu \nu \lambda}  \tilde{\beta}^T_{\mu} \tilde{K} \d_{\nu} \tilde{\beta}_\lambda - i \tilde{\beta}^T_{\mu} \tilde{J}_{\mu} - \frac{i}{4} A_{\mu} j^{\rho}_{\mu} 
\label{eq:Ltopo3}\eeq
with $\tilde{K} = \left(\begin{array}{cc} 8 & 0\\0 & -8\end{array}\right)$ and $\tilde{J} = (j_\rho, \tilde{j}_\theta)$. We see that the $e^{i \varphi_\rho}$ sector is described by a $U(1)_8$ CS theory, while the $e^{i \theta}$ sector  - by the 
$U(1)_{-8}$ CS theory, and the two sectors are decoupled in the Abelian part of the theory. Thus, we can think of the surface topological order (\ref{Eq:TopoLargerho}), $T_{96}$, as embedded into the $\mathrm{Ising}\times U(1)_8 \times U(1)_{-8}$ theory. The subset of the anyons of the latter theory allowed on the surface, coincides precisely with the anyons, which are local with respect to the electron  $c = f e^{4 i \phi} = f e^{4 i \varphi_\rho} e^{4 i \theta}$. 

We note that in the absence of time-reversal symmetry, the topological order $T_{96}$ can certainly be realized in a strictly 2d fermion system (i.e. it is a legal topological order). Further, one can implement this topological order in 2d keeping the global particle-number symmetry and the anyon charge assignments (\ref{Eq:TopoLargerho}). Indeed, begin by forming a layer with $\mathrm{Ising} \times U(1)_{-8}$ topological order built out of electrically neutral, bosonic degrees of freedom. Label the anyons in this ``bosonic" topological order as $\{1, \sigma, f\}  e^{i k \theta}$. Next, make another layer, where electrons are bound into charge $2$ molecules and these molecules form a bosonic $\nu = 1/8$ Laughlin state. This gives a $U(1)_{8}$ topological order with anyons $e^{i m \varphi_\rho}$ carrying electric charge $m/4$. The Hall conductivity of this layer is $\sigma_{xy} = \frac{1}{8} \frac{(2 e)^2}{h} =  \frac12 \frac{e^2}{h}$. 
The topological order of the two layers taken together is $\mathrm{Ising}\times U(1)_{8} \times U(1)_{-8} \times \{1,c\}$. The last factor in the above product encodes the fact that we are dealing with a system made out of electrons $c$. Now, condense the  (bosonic) anyon $e^{4 i \varphi_\rho} e^{4 i \theta} f c$. This identifies $c \sim f e^{4 i \varphi_\rho} e^{4 i \theta}$ and confines all the anyons in $\mathrm{Ising} \times U(1)_8 \times U(1)_{-8}$ that have non-trivial mutual statistics with $f e^{4 i \varphi_\rho} e^{4 i \theta}$. The result is a phase whose intrinsic topological order and anyon charges are identical to that of the ETI surface. However, unlike our surface phase, which is time-reversal invariant and so has an electrical Hall conductivity $\sigma_{xy} = 0$ and thermal Hall conductivity $\kappa_{xy}/T = 0$, the present 2d state has $\sigma_{xy} = 1/2$ and $\kappa_{xy}/T = 1/2$, and so necessarily breaks the time-reversal symmetry. In fact, we can identify this 2d state with an ETI slab, where the top surface is in the symmetry-respecting topologically-ordered phase and the bottom surface is in the topologically trivial $\sigma_{xy} = 1/2$ phase. This identification yields the properties of the edge between the topologically-ordered surface phase and the $\sigma_{xy} = -1/2$ surface phase. The edge carries a right-moving neutral Majorana $(c = 1/2)$ mode $f$, a right-moving bosonic $(c = 1)$ charge mode $\varphi_\rho$ and a left-moving bosonic $(c = -1)$ neutral mode $\theta$.  
The edge action is given by,
\bea L &=&  f (\d_\tau - i v_f \d_x) f  + \frac{8i}{4\pi} \d_{\tau} \varphi_\rho \d_x \varphi_\rho - \frac{8 i}{4 \pi} \d_\tau \theta \d_x \theta  - \frac{i}{\pi} \epsilon_{\mu \nu} A_{\mu} \d_{\nu} \varphi_\rho \nn\\
 &+& V_{\rho\rho} (\d_x \varphi_\rho)^2 + V_{\theta \theta} (\d_x \theta)^2 + V_{\rho \theta} \d_x\theta \d_x \varphi_\rho \label{eq:Ledge}\eea
The edge carries an overall electrical conductance $G = 1/2$ and chiral central charge $c_- = 1/2$, consistent with the difference of electrical and thermal Hall conductances of the  topologically-ordered surface phase and the $\sigma_{xy} = -1/2$ surface phase, which it separates.

We have shown that the surface topological order $T_{96}$ can be realized strictly in two dimensions if we give up the time-reversal symmetry. In section \ref{sec:topoSC} we will also argue that it can be realized strictly in 2d if we give up the particle-number symmetry, but keep the time-reversal symmetry. We will also prove in section \ref{sec:Witten} that there is no 2d realization, which preserves both of these symmetries. However, first we discuss how the time-reversal symmetry is implemented in the topologically-ordered surface phase.

\subsection{Time-reversal symmetry.}
\label{sec:Timp}
We have already discussed how the particle-number symmetry is implemented in the topologically-ordered surface state: the $U(1)_8$ sector carries the electric charge, while the $\mathrm{Ising}$ and $U(1)_{-8}$ sectors are electrically neutral. Now, let us discuss how the time-reversal symmetry is implemented. We take the charge $1/4$ boson, $e^{i \phi}$, to transform trivially into itself under ${\cal T}$. As for the anyons descendant from the vortex excitations of the superconductor, they keep their transformation properties (\ref{eq:Tvort}). One subtlety is that  the descendant of the flux $4 \pi$ vortex, $e^{4 i \theta}$, now transforms into itself, ${\cal T}: e^{4 i \theta} \to e^{-4 i \theta} \sim e^{4 i \theta}$. In principle, this anyon can become either a Kramers singlet or a Kramers doublet. As we show in Appendix \ref{app:4Kramers}, both options are allowed and correspond to two distinct surface states, separated by a surface phase transition. Here, for simplicity, we take $e^{4 i \theta}$ to be a Kramers singlet. Thus, the full transformation properties become:

\bea {\cal T}:\quad &&\sigma e^{i m \phi} e^{i k \theta} \to \sigma e^{i m \phi} e^{-i k \theta}, \,\,\,\,\quad\quad\quad \quad\quad\quad\quad\quad\quad \quad\quad\quad k = \pm 1, \pm 3 \nn\\
&& e^{i m \phi} e^{i k \theta} \to e^{i m \phi} e^{i k\theta} , \quad \,\,\,\,\,\,\,\,{\cal T}^2 = +1,   \quad\quad\quad\quad\quad\quad\quad \,\,\,\, k = 0, 4 \nn\\
&& f e^{i m \phi} e^{i k \theta} \to f e^{i m \phi} e^{ i k \theta} , \quad\, {\cal T}^2 = -1,   \quad\quad\quad\quad\quad \quad\quad\,\,\,\,k = 0, 4 \nn\\
&& e^{i m \phi} e^{i k \theta} \to f e^{i m \phi} e^{-i k \theta}, \quad f e^{i m \phi} e^{i k \theta} \to e^{i m \phi} e^{-i k \theta}, \quad\quad\,\, k = \pm 2  \nn\\
\label{eq:TT96}\eea
For anyons, which transform into themselves under time-reversal, we've indicated whether the anyon is a Kramers singlet or a Kramers doublet. Note that the descendant 
of the Bogolioubov quasiparticle, $f_\sigma$, is naturally a Kramers doublet, as is the electron $e^{4 i \phi} f_\sigma$. The transformation rules (\ref{eq:TT96}) can be easily rewritten in the
$e^{i \varphi_\rho}$, $e^{i \theta}$ basis.

It is easy to see that the time-reversal transformations above are consistent with fusion and braiding rules. Indeed, the effective action in the topologically-ordered phase differs from the flux-tube action for the superconducting surface only by the second term in Eq.~(\ref{eq:Ltopo1}). Now, under ${\cal T}$, $e^{i \phi} \to e^{i \phi}$, while the vortex descendants transform as $a e^{i k \theta} \to b e^{-i k \theta}$. This means that the currents $j_{\phi}$ and $j_\theta$ in Eq.~(\ref{eq:Ltopo1})  transform oppositely under time-reversal. Thus, letting the CS gauge-fields $\alpha_1$ and $\alpha_2$ transform oppositely, we see that the second line in Eq.~(\ref{eq:Ltopo1}) is manifestly time-reversal invariant. Moreover, we already checked the ``emergent" time-reversal symmetry of the flux-tube action for the superconducting surface (first line in Eq.~(\ref{eq:Ltopo1})). Hence, the surface topological order obtained is time-reversal invariant.

\subsection{Back to the superconductor.}
\label{sec:phicond}
In this section we show that starting with the symmetry-preserving topologically-ordered surface described above, we can drive a surface phase transition back to the superconducting phase. The statistics of flux-tubes in this superconductor exactly match those described in section \ref{sec:superfluid}. This justifies the procedure we used in section \ref{sec:cond8pi} for condensing the flux $8\pi$ vortex.

The transition from the topologically-ordered phase to the superconductor is driven by condensing the charge $1/4$ boson $e^{i \phi}$. This anyon transforms trivially under ${\cal T}$, hence its condensation does not break ${\cal T}$. However, since it carries charge, the resulting phase will be a superconductor. The physical, local, Cooper pair order parameter $c c$ is identified with the $8$'th power of $e^{i \phi}$.

Note that $e^{i \phi}$ braids non-trivially with all the flux-tube descendants $a e^{i k \theta}$, $a \in \{1, \sigma, f\}$ with $k \neq 0$. Thus, in the absence of an external gauge field, all the anyons $a e^{i k \theta}$ with $k \neq 0$ will be confined by the condensation. (We will shortly see that these excitations correspond to superconducting vortices, so their confinement is actually only logarithmic). The only stable truly deconfined excitation will be the $f$-fermion, which after the condensation becomes identified with the electron, $c = f e^{4 i \phi} \sim f$. Hence, the resulting superconducting state carries no intrinsic topological order. 

Next, let's gauge the global $U(1)$ symmetry. Now, upon the condensation of $e^{i \phi}$, the anyons $a e^{i k \theta}$ will bind a finite flux $\Phi = \pi k$ such that the Aharonov-Bohm phase $e^{i \Phi/4}$ picked up by $e^{i \phi}$ upon going around $a e^{i k \theta}$ compensates the mutual statistics $e^{-i \pi k /4}$ between $e^{i \phi}$ and $a e^{i k \theta}$. The resulting flux-tubes will be deconfined. Note that since $a e^{i k \theta}$ is electrically neutral, the attached magnetic flux does not alter its statistics. Recall that in section \ref{sec:cond8pi}, we chose the $a e^{i k \theta}$ anyons to have the same statistics as the flux-tubes in the superconductor. 
Thus, all the properties of the $e^{i \phi}$-condensed phase exactly match those of the superconducting surface. The above arguments are formalized in Appendix \ref{app:partvort} using the standard particle-vortex duality.

We expect the transition between the topologically-ordered and superconducting surface phases to be in the $XY^*$ universality class. The field $\psi$ of the $XY$ field-theory is just $\psi = e^{i \phi}$. The star in $XY^*$ serves to remind that the physical Cooper pair order parameter corresponds to the operator $\psi^8$. Note that here we are assuming that the phase transition occurs at fixed surface electron density.

Since the topologically-ordered and superconducting surface phases are separated by a continuous transition in the $XY^*$ universality class, we expect that the edge between them can be gapped out. Indeed, imagine that the $XY^*$ transition is driven by changing a parameter $g$ in the Hamiltonian. Now, slowly tune $g$ as a function of e.g. the $\mathrm{x}$ coordinate on the surface, interpolating between the two phases. Since the low-energy physics can be described by the $XY$ field-theory, and since conventional bosonic superfluids (or $XY$ magnets) generally possess no gapless edge states, we expect a gapped interface.

Now recall that the interface between the topologically-ordered and $\sigma_{xy} = -1/2$ surface states is gapless and described by the action (\ref{eq:Ledge}). Now imagine condensing the $e^{i \phi} = e^{i \varphi_\rho} e^{i \theta}$ anyon in the topologically-ordered region, driving a phase transition to the superconductor. What happens to the edge? The perturbation $\cos(\phi) = \cos(\varphi_\rho + \theta)$ now becomes allowed in the edge theory (\ref{eq:Ledge}). This perturbation is capable of gapping out the $\varphi_\rho, \theta$ modes, leaving the chiral Majorana mode $f$. Thus, we recover the familiar edge between the superconducting and $\sigma_{xy} = -1/2$ surface phases. 
 
\subsection{Reduced topological order.}
\label{sec:reduced}
The shear number of anyons in the surface topologically-ordered state $T_{96}$ that we've constructed is displeasing. In this section, we show that one can drive a surface phase transition out of $T_{96}$ into a different topologically-ordered state, which preserves the time-reversal and particle number conservation symmetries, but has fewer anyons. 

The phase transition is driven by condensing the bosonic anyon $e^{4 i \theta}$. This anyon is neutral and transforms trivially under time-reversal symmetry. Thus, its condensation does not break the global symmetries. In the condensed phase, all the anyons that have nontrivial mutual statistics with $e^{4 i \theta}$ will be confined, and all the excitations differing by $e^{4 i \theta}$ will be identified, leaving,
\bea &&\sigma e^{i m \varphi_\rho} e^{i k \theta}, \quad\quad \quad \quad \quad \quad \,\,\,\,\,\,\,\,\,\, m = 1, 3, 5, 7,\,\, k = 0, 2  \nn\\
&& e^{i m \varphi_\rho} e^{i k \theta},\,\, fe^{i m \varphi_\rho} e^{i k \theta}, \quad \quad\quad  m = 0, 2, 4, 6,\,\, k = 0, 2
\label{Eq:Toposmall} \eea
We see that in the $e^{i \theta}$ sector, only the anti-semion $\bar{s} = e^{2 i \theta}$ survives the phase transition: the condensation reduces $U(1)_{-8} \to U(1)_{-2}$, where $U(1)_{-2} = \{1, \bar{s}\}$. Thus,
the topological order in the condensed phase can be thought of as a subset of the $\mathrm{Ising} \times U(1)_8 \times U(1)_{-2}$ theory. In fact, it is precisely the subset which is local with respect to the electron,  $c = f e^{4 i \varphi_\rho} e^{4 i \theta} \to f e^{4 i \varphi_\rho}$.  Furthermore, we recall that the ordinary Moore-Read state is given by the subset of the $\mathrm{Ising} \times U(1)_8$ theory which is local with respect to the electron $f  e^{4 i \varphi_\rho}$. Thus, the intrinsic topological order in the condensed phase is identical to $\mathrm{Moore-Read} \times U(1)_{-2}$. This state has twice as many anyons as the Moore-Read state: $24$, if we count the electron, and 
$12$, if we don't.

In fact, the charge quantum numbers in the condensed phase are also identical to those of $\mathrm{Moore-Read} \times U(1)_{-2}$ (with the antisemion $\bar{s} = e^{2 i \theta}$ being electrically neutral). As for the time-reversal symmetry, its action can be deduced directly from the transformation properties in the $T_{96}$ phase (\ref{eq:TT96}), and is given in Eq.~(\ref{eq:TT24}). 
Note that although the topological content is a direct product, the time-reversal transformations mix the $\mathrm{Moore-Read}$ and $U(1)_{-2}$ parts. 

We observe that the reduced topologically-ordered phase $\mathrm{Moore-Read}\times U(1)_{-2}$ has no neutral bosons that can be condensed  to further
reduce the number of anyons, while preserving the particle-number symmetry. It is not currently clear if the $\mathrm{Moore-Read} \times U(1)_{-2}$  
topological order is the ``minimal" (in terms of e.g the number of anyons or the total quantum dimension) symmetry-respecting surface termination of an ETI.

We have described a route where the $\mathrm{Moore-Read} \times U(1)_{-2}$ phase is obtained from the surface superconductor  by first condensing the flux $8 \pi$ vortex $e^{8 i \theta}$
and then condensing the anyon $e^{4 i \theta}$, descendant from the flux $4 \pi$ vortex. One can, alternatively, go directly from the surface superconductor to the $\mathrm{Moore-Read} \times U(1)_{-2}$ phase
by condensing the flux $4 \pi$ vortex $e^{4 i \theta}$. The reverse transition from the $\mathrm{Moore-Read} \times U(1)_{-2}$ state to the superconductor is obtained by condensing the charge $1/2$ boson,
$e^{2 i \phi} = e^{2 i \varphi_\rho} \bar{s}$. This confines all the anyons, except for $f$, which becomes identified with the electron $c = e^{4 i \varphi_\rho} f \sim f$. The phase transition is again in the 
$XY^*$ universality class, but the Cooper pair order parameter is now given by the fourth power of the $XY$-field.

Note that the edge between the $\mathrm{Moore-Read} \times U(1)_{-2}$  and the $\sigma_{xy} = -1/2$ surface phases still has the structure (\ref{eq:Ledge}). 
The edge between the $\mathrm{Moore-Read} \times U(1)_{-2}$ phase and the superconducting phase is generally gapped.

\subsection{Witten effect.}
\label{sec:Witten}
In this section we show that states with the same topological content and charge quantum numbers as the $T_{96}$ and $\mathrm{Moore - Read}\times U(1)_{-2}$ phases discussed above cannot be realized strictly in 2d
without breaking the time-reversal symmetry. We then demonstrate that time-reversal invariant realizations of these phases on the surface of an ETI are consistent with the Witten effect in the ETI bulk.

Given a lattice system with particle-number symmetry, we can always couple it to a weakly-fluctuating compact electromagnetic gauge field $A_{\mu}$. For a 2d lattice system, we can then consider instanton events in the coupled
theory where the magnetic flux through the plane changes by $2 \pi$. During such an event, flux $2\pi$ is nucleated through a single plaquette of the lattice (say at the origin) and then allowed to expand to a smooth distribution. The instanton event
is a local process. Thus, it cannot affect the Berry's phase picked up by a distant quasiparticle upon encircling the origin. However, the instanton event clearly changes the Aharonov-Bohm phase picked up by 
a charge $Q$ quasiparticle by $e^{2 \pi i Q}$. Therefore, during the insanton event, an anyon must be excited whose mutual statistics with the charge $Q$ quasiparticle compensates this Aharonov-Bohm phase. More precisely,
the instanton must nucleate an anyon $a$, such that for any anyon $b$, $e^{2 \pi i Q_b} M^{ab} = 1$, where $M^{ab}$ is the mutual statistics of anyons $a$ and $b$. 

Now suppose that the $T_{96}$ phase (with its anyon charge assignments) could be realized strictly in 2d. Applying the above reasoning, we conclude that one of the anyons $e^{-2 i \varphi_\rho}$ or $f e^{2 i \varphi_\rho} e^{4 i \theta}$ has to be nucleated during an instanton event, which changes the magnetic flux by $2 \pi$. These anyons differ by an electron $c = f e^{4 i \varphi_\rho} e^{4 i \theta}$ and carry electric charge $Q = \mp 1/2$. But an instanton event must preserve the electric charge! Now, in general, it is possible that the compensating electric charge is present as a Hall polarization charge, $Q_H = \sigma_{xy}$, carried by the nucleated flux $2\pi$. However, if we assume that the system is time-reversal invariant, the Hall conductivity $\sigma_{xy}$ must vanish. Thus, the phase $T_{96}$ cannot be realized in 2d with both time-reversal and particle-number symmetries preserved. The same holds for the $\mathrm{Moore-Read}\times U(1)_{-2}$ phase, where an identical argument shows that the charge $\mp 1/2$ anyons $e^{-2 i \varphi_\rho}$ or $f e^{2 i \varphi_\rho}$ must be created during an instanton event.

How is the above paradox resolved when the $T_{96}$ (or the $\mathrm{Moore - Read}\times U(1)_{-2}$ phase) is realized on the surface of an ETI? To answer this question we have to recall that the electromagnetic response of an ETI contains a $\theta$-term (\ref{eq:Stheta}), with the $\theta$-angle, $\theta = \pi$.\cite{SCZ}  For a 3d insulator with no intrinsic topological order and a finite $\theta$-angle, a magnetic monopole with flux $2 \pi m$ in the bulk of the insulator acquires an electric charge $Q = n + \frac{\theta m}{2 \pi}$, with $n$ - an integer. This phenomenon is known as the Witten effect.\cite{Witten} The fractional part of the monopole electric charge is fixed by $\theta$, while the integer part $n$ corresponds to the freedom of adding electron excitations on top of the monopole. Thus, magnetic monopoles in the bulk of an ETI carry a half-odd-integer electric charge. In contrast, magnetic monopoles in vacuum carry an integer electric charge. 

An event where the magnetic flux through the ETI surface changes by $2 \pi$ corresponds to a magnetic monopole tunneling through the surface. Let us start with a neutral magnetic monopole in vacuum and let it tunnel through the ETI surface, acquiring an electric charge $1/2$ in the process. There must be a compensating electric charge $-1/2$ left on the surface. If the surface is in the $T_{96}$ phase, the monopole leaves behind the anyon $e^{-2 i \varphi_\rho}$ as it passes through, which carries the required charge $-1/2$. On the other hand, if the monopole acquires an electric charge $-1/2$ as it tunnels through the surface, it leaves behind the $f e^{2 i \varphi_\rho} e^{4 i \theta}$ anyon with charge $1/2$. The excited anyons in the two cases differ by a physical electron, as do the monopoles in the ETI bulk. We conclude that the surface topological order that we've deduced is fully consistent with the Witten effect in the bulk of an ETI.

\subsection{Topological - superconducting slab.}
\label{sec:topoSC}
We have argued in the previous section that the topologically-ordered surface states of an ETI that we've constructed cannot be realized strictly in 2d preservering both time-reversal
and particle-number symmetry. However, if we allow ourselves to break either of these symmetries, this state must be realizable in 2d. In section \ref{sec:cond8pi} we've already discussed
the realization, which breaks the time-reversal symmetry but preserves the $U(1)$ symmetry. This realization corresponds to the ETI slab with the topologically-ordered state on the top surface
and the $\sigma_{xy} = 1/2$ phase on the bottom surface. (We refer to such a system as a topo-$M_+$ slab below). On the other hand, a realization, which preserves the time-revesal symmetry, but breaks particle-number symmetry must be provided
by an ETI slab with the topologically-ordered phase on the top surface and the superconducting phase on the bottom surface. (We refer to such a system as a topo-SC slab below). In this section, we provide additional arguments that a time-reversal invariant 
2d state with the topological content (\ref{Eq:TopoLarge}) and transformation properties (\ref{eq:TT96}) under ${\cal T}$ can, indeed, exist.

Any system of fermions possesses the fermion parity symmetry. By gauging this symmetry one can obtain a bosonic system. Given a 2d topologically-orderered state of fermions it is useful to gauge the fermion
parity symmetry to obtain a topologically-ordered state of bosons. Such an ``extended"  topological order carries more information about the original fermion system than the anyon content and the braiding
rules of the ``fermionic" topological order do. Indeed, even topologically trivial fermion systems, which are equivalent to $n$ copies of the $p_x+ip_y$ superconductor, upon gauging the fermion parity symmetry give rise to $16$ distinct types of bosonic topological order.\cite{KitaevHoneycomb} Further, any time-reversal invariant fermion system must give rise to a time-reversal invariant bosonic topological order upon gauging the fermion parity. This condition may be used to rule out time-reversal invariant strictly 2d implementations of some fermionic topological orders, whose fusion and braiding rules are consistent with ${\cal T}$.\cite{LukaszTSC}

Let us argue that the topo-SC slab introduced above gives rise to a time-reversal invariant bosonic topological order upon gauging fermion parity. First, we need to deduce this bosonic topological order. To do so, it is convenient to
first start with the ${\cal T}$-breaking topo-$M_+$ slab and gauge the fermion parity symmetry. For concreteness, we work here with the ``large" topological order $T_{96}$ (the generalization to the  $\mathrm{Moore-Read}\times U(1)_{-2}$ order can be obtained trivially). The gauged theory must contain an additional anyon corresponding to an electromagnetic flux, $\Phi=\pi$, piercing
the system. The electron must have mutual statistics $-1$ with this anyon. Further, since magnetic flux $2 \pi$ is invisible, 2 $\pi$-fluxes must fuse to an anyon, which exists already in the ungauged fermionic system. For a system with particle-number symmetry, 
we can think of the $\pi$-flux as smeared out over a large area. We call such a smeared $\pi$-flux excitation - the ``elementary" $\pi$-flux. The elemenatry $\pi$-flux must then carry electic charge $Q = \sigma_{xy}/2$. Further, any anyon with charge $Q$ must have mutual statistics
$e^{i \pi Q}$ with the elementary $\pi$-flux.  We can  deduce the statistics of the elementary $\pi$-flux from the effective action for the electromagnetic gauge-field $A_{\mu}$ in the same manner we did in section \ref{sec:superfluid}. The effective action for $A_{\mu}$ is simply given by the Chern-Simons theory
\beq L = -\frac{i k}{4\pi} \epsilon_{\mu \nu \lambda} A_{\mu} \d_{\nu} A_{\lambda} \eeq
with the level $k$ given by the Hall conductivity, $k = \sigma_{xy}$. For the topo-$M_+$ slab, $k = \sigma_{xy} = 1/2$. Writing the current of elementary $\pi$-flux defects as $j^v_{\mu} =\frac{1}{\pi} \epsilon_{\mu \nu \lambda} \d_{\nu} A_{\lambda}$ and repeating the procedure in section \ref{sec:superfluid}, we obtain an effective action for the $\pi$-fluxes
\beq  L = \frac{i}{\pi k} \epsilon_{\mu \nu \lambda} a_{\mu} \d_{\nu} a_{\lambda} - i a_{\mu} j^v_{\mu}\eeq
from which we conclude that the elementary $\pi$ fluxes are Abelian anyons with statistics $e^{\pi i k/4}$. Hence, $\pi$-fluxes through the topo-$M_+$ slab have statistics $e^{\pi i /8}$. 

With the above observations, we can readily guess the ``gauged" topological order for the topo-$M_+$ slab. Recall that the topological order $T_{96}$ corresponds to a subset of the $\mathrm{Ising}\times U(1)_8 \times U(1)_{-8}$ theory given in Eq.~(\ref{Eq:TopoLarge}). The gauged  order corresponds to the full $\mathrm{Ising}\times U(1)_8 \times U(1)_{-8}$ order. The elementary $\pi$-flux is identified with the $e^{i \varphi_\rho}$ anyon, which, indeed, has electric charge $Q = 1/4$, self-statistics $e^{i \pi/8}$ and mutual statistics $e^{i \pi Q}$ with any charge $Q$ anyon. Two $\pi$-fluxes $e^{i \varphi_\rho}$ fuse to $e^{2 i \varphi_\rho}$, which is an allowed excitation in the original $T_{96}$ theory, as required. All the anyons in the full $\mathrm{Ising} \times U(1)_8 \times U(1)_{-8}$ either belong to the $T_{96}$ subtheory or can be obtained by fusing an anyon in $T_{96}$ with the elemenatary $\pi$-flux $e^{i \varphi_\rho}$. 

Having understood the gauged topo-$M_+$ slab, we proceed to the topo-SC slab. In the absence of any symmetry, the topo-SC slab can be obtained from the topo-$M_+$ slab by gluing to its bottom surface an additional ETI slab with the $\sigma_{xy} = -1/2$ state on top and the SC state on the bottom. We call this latter system the $M_-$-SC slab. The Hall-conductivities $\sigma_{xy} = 1/2$ from the bottom surface of the topo-$M_+$ slab and $\sigma_{xy}= -1/2$ from the top surface of the $M_-$-SC slab cancel, so a $\pi$-flux piercing both slabs is only sensitive to the bottom SC surface. Now, the $M_-$ - SC slab is identical to a $p_x-ip_y$ superconductor by the argument of section \ref{sec:superfluid}. Thus, the topo-SC slab can be obtained from the topo-$M_+$ slab by gluing on a $p_x-ip_y$ superconductor. We know that gauging fermion parity in a $p_x-ip_y$ superconductor gives rise to an $\overline{\mathrm{Ising}}$ topological order (i.e. a time-reversal conjugate of $\mathrm{Ising}$), with anyons $\{1, \bar{\sigma}, \bar{f}\}$. Here, $\bar{f}$ corresponds to the electron of the $p_x-ip_y$ superconductor and $\bar{\sigma}$ is the $\pi$-flux. Now, to glue the topo-$M_+$ and $p_x-ip_y$ slabs we need to identify the electron operators in the two theories, $f e^{4 i \varphi_\rho} e^{4 i \theta} \sim \bar{f}$. This is done by condensing the (bosonic) anyon $f \bar{f} e^{4 i \varphi_\rho} e^{4 i \theta}$. Thus, the gauged topo-SC slab is described by an $\mathrm{Ising} \times \overline{\mathrm{Ising}} \times U(1)_8 \times U(1)_{-8}$ theory with $f \bar{f} e^{4 i \varphi_\rho} e^{4 i \theta}$ anyon condensed. We will refer to this theory as  $\mathrm{Ising} \times \overline{\mathrm{Ising}} \times U(1)_8 \times U(1)_{-8}/f \bar{f}  e^{4 i \varphi_\rho} e^{4 i \theta}$. The condensation confines some of the anyons and identifies others, giving rise to the following anyon content,
\bea &&\sigma e^{i m \varphi_\rho} e^{i k \theta}, \quad\quad \quad \quad \quad \quad \,\,\,\,\,\,\,\,\,\, k + m - \mathrm{odd},\,\, 0 \leq m \leq 7,\,\,0 \leq k \leq  7,\,\, \quad\quad  \label{eq:sigma}\\
&& e^{i m \varphi_\rho} e^{i k \theta},\,\, fe^{i m \varphi_\rho} e^{i k \theta}, \quad \quad\quad k + m - \mathrm{even},\,\, 0 \leq m \leq 7,\,\, 0 \leq k \leq 7, \quad \quad \label{eq:1f}\\
&&\bar{\sigma} e^{i m \varphi_\rho} e^{i k \theta}, \quad\quad \quad \quad \quad \quad \,\,\,\,\,\,\,\,\,\, k + m - \mathrm{odd},\,\, 0 \leq m \leq 7,\,\,0 \leq k \leq  7,\,\, \quad\quad   \label{eq:barsigma}\\
&&\sigma \bar{\sigma} e^{i m \varphi_\rho} e^{i k \theta}, \quad\quad \quad \quad \quad \quad \,\,\,\,\,\,  m = 0, 2,\,\, k  = 0 , 2, 4, 6,\quad {\mathrm{and}}\quad m = 1, 3, \,\, k = 1, 3, 5, 7 \quad\quad  \nn\\
\label{eq:sigmabarsigma} \eea
The anyons in (\ref{eq:sigma}), (\ref{eq:1f}) form the initial fermionic $T_{96}$ topological order. They are local with respect to the electron operator $c = f e^{4 i \varphi_\rho} e^{4 i \theta}$. The anyons in (\ref{eq:barsigma}), (\ref{eq:sigmabarsigma}) are the $\pi$-fluxes: they possess mutual statistics $-1$ with the electron, as required. Note that anyons in Eq.~(\ref{eq:sigmabarsigma}) are identified as, $(m, k) \sim (m + 4, k+4)$.  Thus, the gauged topo-SC slab has $144$ anyons. 

So far, we've only identified the intrinsic topological order of the gauged topo-SC slab. We also need to specify the transformation properties under time-reversal. First, we note that the intrinsic topological order has a chiral central charge $c_- = 0$, so there is no obstruction to implementing time-reversal symmetry from this perspective. 
Next note that we actually explicitely broke the time-reversal symmetry in order to identify the intrinsic topological order, so we need to guess how ${\cal T}$ acts on the gauged theory. We already know how the anyons of the original $T_{96}$ theory transform, thus, we only need to deduce the transformation properties of the $\pi$-fluxes. These are strongly constrained. Clearly, $\pi$-fluxes must transform to $\pi$-fluxes. From considerations of topological spin and quantum dimension, $\bar{\sigma} e^{i \varphi_\rho}$ anyon can transform only into one of the following $\bar{\sigma} e^{i \varphi_\rho}$,  $\bar{\sigma} e^{-i \varphi_\rho}$, $\bar{\sigma} e^{i \varphi_\rho} e^{4 i \theta}$ and $\bar{\sigma} e^{-i \varphi_\rho} e^{4 i \theta}$. The choices $\bar{\sigma} e^{i \varphi_\rho} \to \bar{\sigma} e^{i \varphi_\rho}$, $\bar{\sigma} e^{i \varphi_\rho} \to \bar{\sigma} e^{i \varphi_\rho} e^{4 i \theta}$ are not consistent with the action of ${\cal T}$ in the original $T_{96}$ theory. The choice $ \bar{\sigma} e^{i \varphi_\rho} \to \bar{\sigma} e^{-i \varphi_\rho} e^{4 i \theta}$ can be ruled out as follows. We can construct the $\pi$-flux anyons in the following way. Let's first build an ``elementary" $\pi$-flux, by considering a configuration of the electromagnetic field $A_{\mu}$, where the flux is spread-out on the topo-surface and in the bulk of the ETI slab, but concentrated in the vortex core on the superconducting surface. Since the vortex on the superconducting surface carries a Majorana zero mode, the elementary $\pi$-flux must be identified with one of the $\bar{\sigma} e^{i m \varphi_\rho} e^{i k \theta}$ anyons. All other $\pi$-fluxes can be obtained by taking one of the anyons on the topo-surface and fusing it with the elementary $\pi$-flux. Now, a time-reversal conjugate of the elementary $\pi$-flux has an opposite magnetic field. If we take the elementary $\pi$-flux and its time-reversal partner and slowly fuse them, we can clearly only nucleate excitations on the superconducting surface, since the magnetic field in the bulk and on the topo-surface is smooth. Thus, an elementary $\pi$-flux fused with its time-reversal partner can only give $1$ or the electron $\bar{f}$. If we assume that under ${\cal T}$, $\bar{\sigma} e^{i \varphi_\rho} \to \bar{\sigma} e^{-i \varphi_\rho} e^{4 i \theta}$ we can convince ourselves that no $\bar{\sigma} e^{i k \varphi_\rho} e^{i m \theta}$ anyon fuses with its time reversal partner to give $1$ and $\bar{f}$. Thus, the only consistent choice is $\bar{\sigma} e^{i \varphi_\rho} \to \bar{\sigma} e^{-i \varphi_\rho}$. Putting this together with the transformation properties of the anyons in the original $T_{96}$ theory, we have
\bea {\cal T}:\quad &&\bar{\sigma} e^{i m \varphi_\rho} \to \bar{\sigma} e^{-i m \varphi_\rho}, \,\,\,\,\quad\quad\quad \quad\quad\quad\quad\quad \,\,\, m \,-\, \mathrm{odd} \nn\\
&& e^{i m \varphi_\rho} \to e^{i m \varphi_\rho}, \quad \bar{f} e^{i m \varphi_\rho} \to \bar{f} e^{ i m \varphi_\rho}, \,\,\,\,\,\,\,\quad m \equiv 0\,\, (\mathrm{mod} \,\,4) \nn\\
&& e^{i m \varphi_\rho} \to \bar{f} e^{-i m \varphi_\rho}, \quad \bar{f} e^{i m \varphi_\rho} \to e^{-i m \varphi_\rho}, \quad m \equiv 2\,\, (\mathrm{mod} \,\,4) \label{eq:Tvortrho}\eea

Note that Eq.~(\ref{eq:Tvortrho}) is exactly analogous to the transformation rules (\ref{eq:Tvort}) we derived for flux-tubes on the superconducting surface in section \ref{sec:superfluid}. This is not surprising and, in fact,
we can identify the $\bar{\sigma} e^{i \varphi_\rho}$ flux-tube with an ``elementary" $\pi$-flux vortex. We will return to the transformation laws (\ref{eq:Tvortrho}) below. 

Eq.~(\ref{eq:Tvortrho}) only lists the transformations  of elementary flux-tubes (and their multiples). For completeness, we  list the transformation properties of all the flux-tubes,

\bea &&\bar{\sigma} e^{i m \varphi_\rho} e^{i k \theta} \to \bar{\sigma} e^{i (2 k - m) \varphi_\rho} e^{i k \theta} \quad\quad \quad \quad \quad\quad \,\,\,\,\,\,\,\,\,\, k + m - \mathrm{odd},\,\, 0 \leq m \leq 7,\,\,0 \leq k \leq  7,\,\, \quad\quad \nn \\
&&\sigma \bar{\sigma} e^{i m \varphi_\rho} e^{i k \theta} \to \sigma \bar{\sigma} e^{-i (m +2) \varphi_\rho} e^{-i (k +2) \theta}, \quad \quad \quad \quad   m = 0, 2,\,\, k  = 0 , 2, 4, 6
\nn\\
&&\sigma \bar{\sigma} e^{i m \varphi_\rho} e^{i k \theta} \to \sigma \bar{\sigma} e^{- i m\varphi_\rho} e^{- i k \theta}, \quad  \quad\quad\quad\quad\quad\,\,\,\,\,\,m = 1, 3, \,\, k = 1, 3, 5, 7  
\nn\\
\label{eq:Tallfluxes} 
\eea

We now propose a route to realizing the topological order $\mathrm{Ising} \times \overline{\mathrm{Ising}} \times U(1)_8 \times U(1)_{-8}/f\bar{f} e^{4 i \varphi_\rho} e^{4 i \theta}$ with the transformation laws (\ref{eq:TT96}), (\ref{eq:Tallfluxes}) under time-reversal in an exactly soluble lattice model. The strategy follows recent work by L.~Fidkowski, X.~Chen and A.~Vishwanath.\cite{LukaszTSC} Imagine building a Walker-Wang model\cite{WW} based on the ``restricted" $\mathrm{Ising}\times U(1)_{-8}$  braided tensory category introduced in section \ref{sec:superfluid}. Recall that this tensor category is the subset of $\mathrm{Ising} \times U(1)_{-8}$, $\{1, \sigma, f\} e^{i k \theta}$, which is local with respect to $f e^{4 i \theta}$. We label this tensor category, $T_{12}$, (with the subscript referring to the number of anyons). Since $f e^{4 i \theta}$ braids trivially with all the anyons in $T_{12}$, the category is non-modular. The surface of the Walker-Wang model will support the $T_{12}$ state and the bulk will have deconfined fermion excitations corresponding to $f e^{4 i\theta}$, as well as flux-tubes, with which $f e^{4 i \theta}$ has mutual statistics $-1$. Now a slab of the Walker-Wang model will carry the $T_{12}$ topological order on the top surface and the $\overline{T}_{12}$ topological order on the bottom surface. Let us write the anyons in this $\overline{T}_{12}$ state as a subset of $\overline{\mathrm{Ising}} \times U(1)_8 = \{1, \bar{\sigma}, \bar{f}\} e^{i m \varphi_\rho}$.  On the bottom surface, $\bar{f}e^{4 i \varphi_\rho}$ braids trivially with all the anyons and corresponds to the deconfined bulk fermion. Thus, we identify $f e^{4 i \theta} \sim \bar{f} e^{4 i \varphi_\rho}$. We now conjecture that the topological order of the slab of the Walker-Wang model as a whole is $\mathrm{Ising} \times \overline{\mathrm{Ising}} \times U(1)_8 \times U(1)_{-8}/f\bar{f} e^{4 i \varphi_\rho} e^{4 i \theta}$.  This topological order naturally decomposes into anyons on the top and bottom surfaces and $\pi$-fluxes, which can be obtained by fusing the anyons on the top and bottom surfaces with an ``elementary" $\pi$-flux $e^{i \varphi_\rho} e^{i \theta}$. 

As we discussed in section \ref{sec:superfluid}, there is a natural action of time-reversal symmetry on the $T_{12}$ category given in Eq.~(\ref{eq:Tvort}). It may be possible to use the machinery recently developed in Ref.~\onlinecite{LukaszTSC} to implement this time-reversal symmetry in the Walker-Wang model. For a slab of a Walker-Wang model, both surfaces will naturally have the same implementation of time-reversal symmetry. This is consistent with the implementation of time-reversal symmetry by the $\overline{T}_{12}$ subtheory (\ref{eq:Tvortrho}). Note that according to the transformation properties (\ref{eq:TT96}), the elementary $\pi$-fluxes through the bulk of the Walker-Wang model, $e^{i \varphi_\rho} e^{i \theta}$, should transform trivially under ${\cal T}$. 

We would like to stress that we are not using the Walker-Wang model here to build a 3d electron topological insulator. Rather, we are using it to build a 2d system, whose topological order is identical to that of a gauged topo-SC slab. Since we are ultimately interested in getting a 2d state, one might not need to use the full power of the 3d Walker-Wang model to get the desired result. Rather, it might be sufficient to build a strictly 2d Levin-Wen model based on the $T_{12}$ tensor category. However, it is conceptually useful to present the argument in terms of the Walker-Wang model where the $T_{12}$ and $\overline{T}_{12}$ states live on opposite surfaces of the slab. 

\subsection{$\mathrm{SC}^*$ phase.}
\label{sec:SCstar}
In this section, we discuss the phase obtained from the topologically-ordered, symmetry-preserving ETI surface state by condensing a charge $1$ boson, $e^{4 i\phi} = e^{4 i \varphi_\rho} e^{4 i \theta}$. Since the condensing 
anyon carries electric charge, the resulting phase will be a superconductor. However, the condensation of $e^{4 i \phi}$ does not confine all the anyons, so the resulting superconductor will have a remnant topological order. We, thus, label 
this phase as $\mathrm{SC}^*$. It turns out that the remnant topological order is Abelian. The implementation of time-reversal symmetry in this Abelian phase is still non-trivial. As before, a 2d system realizing the $\mathrm{SC}^*$ topological order (and its implementation of the time-reversal symmetry) must be provided by an ETI slab with the $\mathrm{SC}^*$ phase on the top surface and the ordinary SC phase on the bottom surface. In the present case, we will be able to explicitely construct a 2d model corresponding to such a slab.

To reduce the complexity, in this section we work with the $\mathrm{Moore-Read}\times U(1)_{-2}$ surface topological order (\ref{Eq:Toposmall}). Recall that this topological order consists of the subset of the $\mathrm{Ising}\times U(1)_8 \times U(1)_{-2}$ theory, which is local with respect to the electron $f e^{4 i \varphi_\rho}$. Now, condensation of the charge $1$ boson $e^{4 i \phi} = e^{4 i \varphi_\rho}$ leaves the following anyons, $\{1, e^{2 i\varphi_\rho}\}\times\{1, \bar{s}\}\times\{1, f\}$. Note that $f$ now becomes identified with the electron. Labeling the semion $e^{2i \varphi_\rho} = s$, the topological order becomes equivalent to  $U(1)_2 \times U(1)_{-2}\times \{1,f\}$. There are $8$ anyon types, counting the electron, and $4$ - modulo the electron. 

The transformation properties of the anyons under ${\cal T}$ are inherited from the $\mathrm{Moore-Read}\times U(1)_{-2}$ phase, Eq.~(\ref{eq:TT24}), and read,
\bea && s \bar{s} \to s \bar{s}, \quad {\cal T}^2 = 1\\
&& f \to f, \quad {\cal T}^2 = -1\\
&& f s \bar{s} \to f s \bar{s}, \quad {\cal T}^2 = -1\\
&& s \to sf,\quad sf \to s\\
&& \bar{s} \to \bar{s} f, \quad \bar{s} f \to \bar{s}\eea
We see that the semions $s$ and $\bar{s}$ transform non-trivially under ${\cal T}$. We now construct a 2d system with identical topological order and implementation of time-reversal symmetry. 

We begin with a bosonic topological order consisting of two layers. In the first layer, we form a $U(1)_2 \times U(1)_{-2}$ topological order with anyons labelled as $\{1, u\} \times \{1, \bar{u}\}$. We take
 $U(1)_2$ and $U(1)_{-2}$ sectors to be exact time-reversal conjugates of each other, so that $u \leftrightarrow \bar{u}$ under ${\cal T}$, and $u \bar{u} \to u \bar{u}$ with ${\cal T}^2 = 1$. In the second layer, 
we form a toric code, $Z_2$, consisting of anyons $\{1, e, m, \epsilon \}$. As usual, $e$ and $m$ are self-bosons and mutual semions, while $\epsilon$ is a fermion. We let these anyons transform under ${\cal T}$
as, $e \leftrightarrow m$ and $\epsilon \to \epsilon$ with ${\cal T}^2 = -1$. It is known that such a toric code can be obtained by starting with a non-interacting 2d superconductor in the ${\bf DIII}$ universality class
and gauging the fermion parity symmetry.\cite{LukaszTSC} The edge of this unusual toric code carries two counter-propagating Majorana modes.

So far, we've constructed a bosonic topological order $U(1)_2 \times U(1)_{-2} \times Z_2$. We now put this system on top of a trivial 2d electron insulator and condense the boson $u \bar{u} \epsilon c$, with $c$ - the physical
electron. Note that this boson is a Kramers singlet. There are 8 deconfined anyons: $\{u \bar{u}, \epsilon, u \bar{u} \epsilon, u e, u m, \bar{u} e, \bar{u} m\}$. The physical electron is now identified with $c \sim u \bar{u} \epsilon$. We see that the resulting topological order and transformation properties under ${\cal T}$ are identical to that of the $\mathrm{SC}^*$ phase on the surface of an ETI, with the correspondence $s \sim u e$, $\bar{s} \sim \bar{u} e$, $f \sim u \bar{u} \epsilon$.

Note that the 2d topological order we've constructed appers to have a gapless edge. Indeed, the edge of $U(1)_2 \times U(1)_{-2}$ can be gapped out by condensing $u \bar{u}$ on the edge. This leaves the gapless edge of the toric code consisting of two counterpropagating
Majorana modes. Note that when $u \bar{u}$ is condensed on the edge, an electron can tunnel into the Majorana mode $\epsilon$, as the electron $c = u \bar{u} \epsilon \sim \epsilon$.  Given the toric code by itself, it is not possible to gap out these counter-propagating Majorana modes without breaking the time-revesal symmetry. It is not immediately clear if this is also the case when the $U(1)_2 \times U(1)_{-2}$ topological order is present (even
though there is a regime where the edge modes corresponding to $U(1)_2 \times U(1)_{-2}$ are gapped, an edge phase transition, or sequence of phase transitions, involving both $U(1)_2 \times U(1)_{-2}$ and toric code sectors, which gaps all the modes out is not immediately ruled out). 

We note that the edge modes can certainly be eliminated by gluing on a 2d topological superconductor in the class ${\bf DIII}$. The edge then has two pairs of counter-propagating Majorana modes, which can be gapped out without breaking ${\cal T}$. In terms of the intrinsic topological order and the transformation properties of anyons under time-reversal, our 2d systems without a ${\bf DIII}$ superconductor glued on and with a ${\bf DIII}$ superconductor glued on are identical. Nevertheless, they correspond to different phases of matter - this can easily be checked by gauging the fermion parity symmetry and verifying that the $\pi$-fluxes transform differently under ${\cal T}$ in the two cases. We call these two phases of matter $\mathrm{SC}^*_{2d,a}$ and $\mathrm{SC}^*_{2d,b}$. 

The above two distinct 2d phases of matter have a natural interpretation in terms of an ETI slab with the $\mathrm{SC}^*$ phase on top and the ordinary $\mathrm{SC}$ phase on the bottom. Indeed, it is well known that a domain wall on the ordinary superconducting surface of an ETI across which the sign of the superconducting order parameter changes carries two counter-propagating Majorana modes.\cite{FuKaneMajorana} Thus, a slab of an ETI with both surfaces in the SC phase, with opposite signs of the order parameter on the two faces is identical to a 2d topological superconductor in the ${\bf DIII}$ universality class. Further, changing the sign of the order parameter on some region of the SC surface is identical to gluing on a ${\bf DIII}$ superconductor to that region. An ETI slab with $\mathrm{SC}^*$ phase on top and the ordinary SC phase on the bottom is equivalent to the $\mathrm{SC}^*_{2d}$ phases, with the particular realization, $a$ vs $b$, depending on whether the sign of the order parameter is the same or different on the two surfaces. 

The discussion in this section of the surface SC$^*$ phase and the SC$^*$-SC slab of an ETI serves as yet another consistency check on our construction of the symmetry-respecting topologically-ordered surface phase.

\section{Conclusion.}
In this paper we have deduced a gapped symmetry-respecting surface phase of an electron topological insulator.  This phase carries an intrinsic Moore-Read$\times U(1)_{-2}$ topological order.  We have argued that this phase can be obtained from the superconducting surface phase through a direct second order phase transition involving proliferation of flux $4 \pi$ vortices. It would be extremely interesting to identify possible microscopic interactions on the ETI surface that may be sympathetic to such vortex condensation. This would be the first step to a long-term goal of realizing such a symmetry-preserving topologically-ordered surface phase experimentally.

The Moore-Read$\times U(1)_{-2}$ topological order supports $24$ anyon types ($12$-modulo the electron) and has a total quantum dimension ${\cal D} = \sqrt{32}$ (${\cal D} = 4$, modulo the electron).  It is currently an open question whether this is the ``smallest" possible gapped, symmetry-preserving surface phase of an ETI ranked e.g. by the number of anyon types or the total quantum dimension. Actually, a different topologically-ordered ETI termination with just $12$ anyons ($6$-modulo the electron) has been proposed in Ref.~\onlinecite{BNQ} and independently in Ref.~\onlinecite{AshvinETI}. In terms of its intrinsic topological order, this phase is identical to the ``restricted" Ising$\times U(1)_{-8}$ braided tensor category (also labeled as $T_{12}$), discussed in sections \ref{sec:superfluid} and \ref{sec:topoSC}, however, the anyons are endowed with electric charge quantum numbers and the time-reversal symmetry is implemented differently. Using an exactly-soluble 3d Walker-Wang model, Ref.~\onlinecite{AshvinETI} has demonstrated that this topological order can appear as a surface termination of some non-trivial ${\cal T}$-invariant 3d electron insulator with $\theta = \pi$. However, it is not clear whether the resulting 3d phase is continuously connected to the standard (non-interacting) ETI. This question is currently under active investigation. If it is, then a surface phase transition (or sequence of phase transitions) between the Moore-Read$\times U(1)_{-2}$ state discussed in this paper and the $T_{12}$ state of Refs.~\onlinecite{BNQ,AshvinETI} must exist. 

We saw that symmetry is implemented in the surface Moore-Read$\times U(1)_{-2}$ phase in a way that is prohibited strictly in 2d. This is believed to be a general property of the boundary states of SPT phases. One may then ask what are the general consistency conditions on implementation of a global symmetry $G$ in a 2d topologically-ordered phase $T$ (note that in a fermionic system, $G$ always includes the fermion parity symmetry). When $G$ is unitary, a necessary requirement is that the topological order $T$ can be consistently extended by ``weakly gauging" the global symmetry $G$.  
However, if $G$ includes the anti-unitary time-reversal symmetry no analogue of the gauging procedure is known. A further question is whether, given an intrinsic topological order $T$ equipped with some action of a global symmetry $G$, which is consistent with the fusion and braiding rules, but nevertheless cannot be implemented strictly in 2d, one can ${\it always}$ realize $T$ as a surface of some 3d SPT phase. We hope that a detailed understanding of  examples, such as the one presented in this paper, might provide a stepping stone to answering these very general questions. 

{\it Note added}: Upon completing this paper, we have learned that an identical topologically-ordered surface termination of an ETI has been concurrently obtained by C.~Wang, A.~C.~Potter and T.~Senthil.\cite{ChongDrewSurface} The same authors have also arrived at the conclusion that electron insulators with $\theta = \pi$ and ${\cal T}^2 = +1$ do not exist.\cite{ChongDrewBulk}

\acknowledgements
We are extremely grateful to P.~Bonderson, C.~Nayak, and T.~Grover for many stimulating discussions. We would particularly like to thank P.~Bonderson, C.~Nayak and X.-L.~Qi for sharing and discussing their unpublished results. We also thank L.~Fidkowski and A.~Vishwanath, and T.~Senthil for sharing their results prior to publication. This research was supported in part by the National Science Foundation under Grant No. NSF PHY11-25915,  
DMR-1101912 (M.P.A.F.) and DMR 0906175 (C.L.K.), by the Caltech Institute of Quantum Information and Matter, an NSF Physics Frontiers Center with support of the Gordon and Betty Moore Foundation (M.P.A.F.), and by a Simons Investigator award from the Simons Foundation (C.L.K.).

\appendix

\section{Mutual semions, which are time-reversal partners.}
\label{app:Kramers}
In this appendix we demonstrate that if two anyons $a$ and $b$ have mutual semion statistics and are mapped into each other by time-reversal symmetry then they must fuse 
to a fermion $c$, which is a Kramers doublet.  Note that if  $a$ and $b$ are Abelian anyons then their fusion product $c$ is unique. On the other hand, if $a$ and $b$ are non-Abelian, 
we assume that we are considering a fusion channel $c$, with $c$-Abelian and $c$ transforming into itself under ${\cal T}$. The fact that $c$ must be a fermion was already explained in section
\ref{sec:superfluid}. We now argue that $c$ must be a Kramers doublet.

Consider two anyons $a$ and $b$ in some region of two-dimensional space. Once we've specified the channel $c$, in which $a$ and $b$ are fused, the only degrees of freedom remaining are
the spatial coordinates $\vec{x}_a$ and $\vec{x}_b$ of $a$ and $b$. We can then write the wave-function of the two-anyon system as $\psi(\vec{x}_a,\ \vec{x}_b)$. The semionic statistical interaction between $a$ and $b$ 
can be implemented by requiring $\psi(\vec{x}_a, \vec{x}_b)$ to have a branch-cut in the $\vec{r} = \vec{x}_a - \vec{x}_b$ plane starting at the origin and running to infinity. The sign of $\psi$ changes
as $\vec{r}$ crosses the branch-cut. Different choices of the branch-cut path are gauge-equivalent, however, once a choice is made, one must work with wave-functions in the corresponding Hilbert space. Here we choose
a branch cut along the positive-$\mathrm{x}$ axis.

We would now like to implement the time-reversal symmetry, which transforms the anyons $a$ and $b$ into each other, i.e. $\vec{x}_a \leftrightarrow \vec{x}_b$. Under a simple complex-conjugation, followed by an interchange of the anyon positions $\psi(\vec{x}_a, \vec{x}_b) \to \psi^*(\vec{x}_b, \vec{x}_a)$, the branch-cut in the $\vec{r} = \vec{x}_a - \vec{x}_b$ plane is transformed to run along the negative-$\mathrm{x}$ axis. To return the branch-cut to its original position, we must follow the above transformation by a gauge-rotation, $\psi(\vec{x}_a, \vec{x}_b) \to \mathrm{sign}(y_a - y_b) \psi(\vec{x}_a, \vec{x}_b)$. The full time-reversal operation then becomes,
\beq {\cal T}: \quad \psi(\vec{x}_a, \vec{x}_b) \to \mathrm{sign}(y_a - y_b) \psi^*(\vec{x}_b, \vec{x}_a)\eeq
and we find ${\cal T}^2 = -1$. Thus, the fusion product of $a$ and $b$ must transform as a Kramers doublet under ${\cal T}$. 

\section{Fermion topological insulators with ${\cal T}^2 = +1$ and $\theta = \pi$ do not exist.}
\label{app:thetapi}
Suppose they did. Let us recall that at a finite $\theta$ angle, a monopole of flux $m$ carries an electric charge $q = n + \frac{\theta m}{2 \pi}$, with $n$ - an arbitrary integer.
At $\theta = \pi$ single monopoles carry a half-odd-integer charge. Under time-reversal an excitation $(q,m)$ with charge $q$ and flux $m$ transforms as $(q, m) \to (q, -m)$. 
In particular, dyons $(1/2, 1)$ and $(1/2, -1)$ are time-reversal partners. These time-reversal partners fuse to the electron, $(1,0)$. 
We would like to show that the electron is a Kramers doublet. 

Recall that the $\theta$ angle does not affect the statistical interaction between dyons. So, to understand the statistical interaction, we may start with $\theta = 0$
and continuously tune $\theta$ to any desired value (e.g. $\theta =\pi$). Thus, the statistical interaction between $(1/2,1)$ and $(1/2, -1)$ at $\theta = \pi$ is the same as the statistical interaction 
between $(0, 1)$ and $(1, -1)$ at $\theta = 0$. Now, the statistical interaction between $(0,1)$ and $(1,-1)$  is actually the same as the statistical interaction between $(0,1)$ and $(1,0)$
(all at $\theta = 0$),  since the monopole-monopole interaction is not statistical. Thus, $(1/2,1)$ and $(1/2, -1)$ interact statistically like a charge and a monopole. 

The statistical interaction between a charge and a monopole is described by the Hamiltonian,\cite{Goldhaber}
\beq H = \frac{1}{2M_q} (\vec{p}_q - \vec{A}(\vec{x}_q - \vec{x}_m))^2 +   \frac{1}{2M_m} (\vec{p}_m + \vec{A}(\vec{x}_q - \vec{x}_m))^2 \label{eq:H}\eeq
 with $\vec{A}(\vec{x})$ - the vector potential of a monopole ($\nabla \times \vec{A}(\vec{x}) = \frac{\hat{r}}{2 r^2}$). One may use a particular
gauge choice,
\beq \vec{A}(\vec{x}) = \frac{1 - \cos \theta}{2 r \sin \theta} \hat{\varphi}\eeq

Here, for concreteness, the non-relativistic limit has been taken; however, the argument does not rely on the specific form of the 
Hamiltonian, but rather on the form of the covariant derivatives. This form may be derived by starting with the Maxwell action with monopole and charge sources, integrating out the gauge-field and then expanding the 
result in charge and monopole velocities. 

According to our discussion above, we may think of the charge and the monopole in Eq.~(\ref{eq:H}), as the $(1/2, -1)$ and $(1/2, 1)$ dyons, respectively. Since these are mapped into each other by time-reversal
symmetry, we set $M_q = M_m$. Note that when the monopole is placed at the origin, a charge moving around a curve ${\cal C}$, picks up a Berry's phase $e^{i \Omega/2}$, where $\Omega$ is the solid
angle subtended by ${\cal C}$. On the other hand, if the charge is placed at the origin, a monople moving around the curve ${\cal C}$, picks up a Berry's phase $e^{-i \Omega/2}$. This is consistent with time-reversal symmetry, which interchanges the charge and monopole, i.e. $(1/2, -1)$ and $(1/2, 1)$. 

Let us now find an explicit expression for an operator which implements the time-reversal symmetry, interchanging the charge and the monopole. Under simple complex conjugation $C:\,\, \psi(\vec{x}_q, \vec{x}_m) \to \psi^*(\vec{x}_q, \vec{x}_m)$, followed by exchange of charge and monopole coordinates, $S:\,\, \psi(\vec{x}_q, \vec{x}_m) \to \psi(\vec{x}_m, \vec{x}_q)$, we find,
\bea ST: \quad &&\vec{p}_q - \vec{A}(\vec{x}_q -\vec{x}_m) \to - \vec{p}_m - \vec{A}(\vec{x}_m - \vec{x}_q) = - \left(\vec{p}_m + \tilde{\vec{A}}(\vec{x}_q - \vec{x}_m)\right)\\
&& \vec{p}_m + \vec{A}(\vec{x}_q -\vec{x}_m) \to - \vec{p}_q + \vec{A}(\vec{x}_m - \vec{x}_q) = - \left(\vec{p}_q - \tilde{\vec{A}}(\vec{x}_q - \vec{x}_m)\right)\eea
with
\beq \tilde{\vec{A}}(\vec{x}) = \vec{A}(-\vec{x}) = \frac{-1 - \cos \theta}{2 r \sin \theta} \hat{\varphi}\eeq
We observe that $\tilde{\vec{A}}$ is just a gauge transform of $\vec{A}$,
\beq \vec{\tilde{A}}(\vec{x}) = \vec{A}(\vec{x}) + \nabla \alpha(\vec{x})\eeq
with
\beq e^{i\alpha(\vec{x})} = e^{-i \varphi}\eeq
Thus, to restore the Hamiltonian (\ref{eq:H}) to its original form, we must perform a gauge transformation, $U :\,\, \psi(\vec{x}_q, \vec{x}_m) \to e^{i \alpha(\vec{x}_q - \vec{x}_m)} \psi(\vec{x}_q, \vec{x}_m)$. Hence,
the form of the time-reversal operator is ${\cal T} = U S C$. It is easy to check that ${\cal T}^2 = -1$. Thus, every bound state of $(1/2, -1)$ and $(1/2, 1)$ dyons at $\theta = \pi$ is a Kramers doublet. Since this bound state is the electron, we arrive at the desired conclusion. We note that our argument is completely insensitive to presence/absence of any rotational/translational symmetry and depends only on the form of the statistical interaction between dyons.


\section{Changing Kramers parity of the $e^{4 i \theta}$ anyon.}
\label{app:4Kramers}
In this appendix we show that there are two possible realization of the $T_{96}$ phase on the surface of the electron topological insulator that differ by the action of ${\cal T}^2$ on the $e^{4 i \theta}$ anyon.
The two phases are separated by a surface phase transition.

First, let us construct a strictly 2d $Z_8$ gauge theory with a global particle-number symmetry and time-reversal. We would like the $Z_8$ charges, $e^{i \tilde{\phi}}$, to carry physical electric charge $1/4$ and transform
trivially under ${\cal T}$ and the $Z_8$ fluxes, $e^{i\tilde{\theta}}$, to be electrically neutral and transform under ${\cal T}$ as $e^{i\tilde{\theta}} \to e^{-i \tilde{\theta}}$. Moreover, we would like the anyon $e^{4 i \theta}$, which
transforms into its own topological sector $e^{4 i \theta} \to e^{-4 i \theta} \equiv e^{4 i \theta}$ to be a Kramers doublet. We now give an explicit construction of this state using the standard Chern-Simons $K$-matrix approach.

The $Z_8$ topological order can be described with a mutual Chern-Simons theory,
\beq L_{Z_8} = \frac{i}{4 \pi} \epsilon_{\mu \nu \lambda}  a^T_{\mu} K \d_{\nu} a_\lambda - i a^T_{\mu} J_{\mu} - \frac{i}{2 \pi} \epsilon_{\mu \nu \lambda} A_{\mu} t^T \d_{\nu} a_\lambda\eeq
where $K = -\left(\begin{array}{cc} 0 & 8\\8 & 0\end{array}\right)$. Here, $a = (a_1, a_2)$ is a pair of CS gauge fields and  $J = (j_{\tilde{\phi}}, j_{\tilde{\theta}})$, is a pair of corresponding currents. $j_{\tilde{\phi}}$ is the current of $Z_8$ charges, $e^{i \tilde{\phi}}$,
and $j_{\tilde{\theta}}$ is the current of $Z_8$ fluxes, $e^{i \tilde{\theta}}$. $t$ is the charge vector, which we take to be $t = (0, 2)$. We are thinking of the $Z_8$ state as a strictly ``bosonic" topological order made out of Cooper pairs,
so entries of $t$ are even integers. The quasiparticles then carry physical electric charges $K^{-1} t = (1/4, 0)$, i.e. $e^{i \tilde{\phi}}$ carries charge $1/4$ and $e^{i \tilde{\theta}}$ is neutral. 
We take the fields to transform under time reversal as,
\bea j^{\tilde{\phi}}_{\mu} \to \lambda_{\mu} j^{\tilde{\phi}}_{\mu}, \quad\quad j^{\tilde{\theta}}_{\mu} \to - \lambda_{\mu} j^{\tilde{\theta}}_{\mu},\\
a^1_{\mu} \to \lambda_{\mu} a^1_{\mu}, \quad\quad a^2_{\mu} \to -\lambda_{\mu} a^{2}_{\mu}\eea
with $\lambda = (1, -1, -1)$. This means that the $Z_8$ charges transform into themselves under time-reversal, while $Z_8$ fluxes transform into anti-fluxes. It is actually more convenient to discuss the time-reversal transformations in the edge theory,
\beq L_{\mathrm{edge}} = \frac{1}{4 \pi} \d_t \chi^T  K \d_x \chi - \d_x \chi^{T} V \d_ x \chi \eeq
with $\chi = (\tilde{\phi}, \tilde{\theta})$. Here, we've switched to real time $t$ from imaginary time $\tau$. We take $\chi$ to transform under time-reversal (in real time) as,
\beq {\cal T}:\,\,\, \tilde{\phi} \to - \tilde{\phi}, \quad\quad \tilde{\theta} \to \tilde{\theta}  + \frac{\pi}{8}\eeq
Since ${\cal T}$ is anti-unitary, this implies,
\beq {\cal T}:\,\,\, e^{i \tilde{\phi}} \to e^{i \tilde{\phi}}, \quad e^{i \tilde{\theta}} \to e^{-i \pi/8} e^{-i \tilde{\theta}}\eeq
Under ${\cal T}^2$,
\beq {\cal T}^2:\quad e^{i k \tilde{\theta}} \to e^{i k \pi/4} e^{i k \tilde{\theta}} \label{eq:T2Z8}\eeq
Note that $e^{8 i \tilde{\theta}}$ transforms trivially under ${\cal T}^2$ as is necessary for an operator in the topologically trivial sector. For $k \neq 4$ in Eq.~(\ref{eq:T2Z8}), the phase factor $e^{i k \pi/4}$ in the transformation properties does not
carry a physical significance, since the time-reversal partners $e^{i k \tilde{\theta}}$ and $e^{-i k \tilde{\theta}}$ are in distinct topological sectors. However, for $k = 4$, $e^{4 i \tilde{\theta}}$ and $e^{-4 i \tilde{\theta}}$ are in the same topological sector. Eq.~(\ref{eq:T2Z8})
then implies that $e^{4 i \tilde{\theta}}$ is a Kramers doublet. Thus, we have constructed a $Z_8$ state with the desired symmetry properties.

Next, imagine starting with the ETI surface in the topologically-ordered $T_{96}$ phase with $e^{4 i \theta}$ being a Kramers singlet. Let's glue on a $Z_8$ phase constructed above onto the surface. Next, condense the anyon $e^{i \phi} e^{-i \tilde{\phi}}$. 
This anyon is electrically neutral and transforms trivially under ${\cal T}$, so the condensation preserves all symmetries. The deconfined anyons are the same as in Eq.~(\ref{Eq:TopoLarge}), except with the replacement $e^{i k \theta} \to e^{i k \theta'} = e^{i k \theta} e^{i k \tilde{\theta}}$. 
The resulting topological order is identical to $T_{96}$. The transformation properties under ${\cal T}$ are also identical, except $e^{4 i \theta'}$ is now a Kramers doublet.


\section{$e^{i \phi}$ condensation in the topologically-ordered surface phase.}
\label{app:partvort}
In this appendix we formalize the argument of section \ref{sec:phicond} using the standard particle-vortex duality. We begin with the effective theory of the topologically-ordered surface, Eq.~(\ref{eq:Ltopo1}). We wish to condense the $e^{i \phi}$ particle. We write the conserved current of $e^{i \phi}$ as $j^\phi_{\mu} = \frac{1}{2\pi} \epsilon_{\mu \nu \lambda} \d_{\nu} b_\lambda$. On a lattice, $b_\mu$ would be a $2 \pi \mathbb{Z}$ valued variable. One can enforce this integer-valued constraint by adding a term  $i b_\mu j^8_\mu$ to the action and summing over all integer values of $j^8_\mu$. Physically, $j^8_{\mu}$ is the current of the vortex in $e^{i \phi}$. As $e^{i \phi}$ carries charge $1/4$, the vortex has flux $\Phi = 8\pi$. We will explicitely confirm this identification below. The phase where $e^{i \phi}$ is condensed corresponds to the worldlines $j^\phi_{\mu}$ proliferating. Correspondingly, the vortex worlines $j^8_{\mu}$ will be supressed. With the above remarks in mind, the effective action becomes,
\bea L_{\mathrm{topo}} &=& L_{\mathrm{Ising}}[j_\theta, j_f]  +  \frac{-8 i}{4\pi} \epsilon_{\mu \nu \lambda} a_{\mu} \d_{\nu} a_{\lambda} - i a_{\mu} j^{\theta}_{\mu} \nn\\
&-& \frac{8 i}{2 \pi} \epsilon_{\mu \nu \lambda}  \alpha^1_{\mu}  \d_{\nu} \alpha^2_\lambda - \frac{i}{2\pi}  \epsilon_{\mu \nu \lambda}  \alpha^1_{\mu}  \d_{\nu} b_\lambda 
- i \alpha^2_{\mu} j^\theta_{\mu} + i b_{\mu} j^8_{\mu}  -\frac{i}{4 \cdot 2 \pi} \epsilon_{\mu \nu \lambda}  A_{\mu}  \d_{\nu} b_\lambda \label{eq:dual1}\eea
We can integrate over $\alpha^1$ in Eq.~(\ref{eq:dual1}). This gives a constraint, $b_{\mu} = - 8 \alpha^2_{\mu}$. The resulting action becomes,
\bea L_{\mathrm{topo}} &=& L_{\mathrm{Ising}}[j_\theta, j_f]  +  \frac{-8 i}{4\pi} \epsilon_{\mu \nu \lambda} a_{\mu} \d_{\nu} a_{\lambda} - i a_{\mu} j^{\theta}_{\mu} \nn\\
&&-i \alpha^2_{\mu} (j^\theta_{\mu} + 8 j^8_{\mu})   + \frac{i}{\pi} \epsilon_{\mu \nu \lambda}  A_{\mu}  \d_{\nu} \alpha^2_\lambda \label{eq:dual2}\eea
Let us discuss the dynamics of the action (\ref{eq:dual2}) in the phase where $e^{i \phi}$ is condensed. As already remarked,  in this phase the fluctuations of $j^8_{\mu}$ will be supressed. We also assume that the anyons $a e^{i k \theta}$ remain gapped
through the phase transition, so the fluctuations of $j^\theta_{\mu}$ are supressed as well. Thus, in the absence of an external electromagnetic field $A_{\mu}$, the gauge field $\alpha^2_\mu$ will be gapless. (The gapless fluctuations will be governed by a kinetic term $(\epsilon_{\mu \nu \lambda} \d_{\nu} \alpha^2_{\lambda})^2$ in the action, which we've dropped in our schematic exposition.).  The corresponding photon is identified with the superfluid Goldstone mode, since  the physical electromagnetic current is given by $J^{EM}_{\mu} = -\frac{1}{\pi} \epsilon_{\mu \nu \lambda} \d_{\nu} \alpha^2_\lambda$. The current $j^{\theta}$ is minimally coupled to $\alpha^2$, while $j^8$ carries charge $8$ under $\alpha^2$. Hence, we identify $j^{\theta}$ with the elementary superconducting vortex and $j^{8}$ with the vorticity $8$ vortex. Thus, $e^{i \phi}$ condensation ``restores" the flux $8 \pi$ vortex as a stable excitation. As long as $A_{\mu}$ is switched off, the superconducting vorticies will be logarithmically confined by the gapless gauge field $\alpha^2$. 

Once a dynamical electromagnetic field $A_{\mu}$ is introduced,
$\alpha^2$ becomes gapped (the Goldstone mode is eliminated). We can then integrate over $\alpha^2$ in Eq.~(\ref{eq:dual2}) to obtain a constraint, $j^\theta_{\mu} + 8 j^8_{\mu} = \frac{1}{\pi} \epsilon_{\mu \nu \lambda}  \d_{\nu} A_\lambda$, i.e. $j^{\theta}$ binds flux $\Phi = \pi$ and $j^{8}$ binds $\Phi = 8 \pi$. The effective ``topological action" for the resultant flux-tubes is now given by the first line in Eq.~(\ref{eq:dual2}), which coincides with the flux-tube action (\ref{eq:Stop2}) obtained in section \ref{sec:superfluid}. Note that the current $j^8$ does not enter this topological part of the action - i.e. the flux $8\pi$ flux-tube is statistically trivial, in agreement with the results of section \ref{sec:superfluid}.


\begin{thebibliography}{}

\bibitem{TI}M.~Z.~Hasan and C.~L.~Kane,  Rev. Mod. Phys. {\bf 82}, 3045 (2010).~ X.-L.~Qi and S.-C.~Zhang,  Rev.~Mod.~Phys. {\bf 83}, 1057 (2011). ~
M.~Z.~Hasan and J.~E.~Moore,  Annu. Rev. Condens. Matter Phys. {\bf 2}, 55 (2011).
\bibitem{FuKaneMele}  L.~Fu, C.~L.~Kane and E.~J.~Mele, Phys. Rev. Lett. {\bf 98}, 106803 (2007).
\bibitem{MooreBalents} J.~E.~Moore and L.~Balents,  Phys. Rev. B {\bf 75}, 121306(R) (2007).
\bibitem{Roy} R.~Roy , Phys. Rev. B {\bf 79}, 195322 (2009).
\bibitem{Hasan} D.~Hsieh et al., Nature (London) {\bf 452}, 970 (2008).

\bibitem{Ludwig}  A.~P.~Schnyder, S.~Ryu, A.~Furusaki, and A.~W.~W.~Ludwig, Phys. Rev. B {\bf 78}, 195125 (2008).
\bibitem{KitaevFree} A.~Kitaev, AIP Conf. Proc. {\bf 1134}, 22 (2009).
\bibitem{Fidkowski1d} L.~Fidkowski and A.~Kitaev, Phys. Rev. B {\bf 81}, 134509 (2010).
\bibitem{Ari} A.~M.~Turner, F.~Pollman, and E.~Berg, Phys.~Rev.~B {\bf 83}, 075102 (2011).
\bibitem{Wen1d} X.~Chen, Z.-C.~ Gu and X.-G.~Wen, Phys. Rev. B {\bf 83}, 035107 (2011).
\bibitem{Cirac} N.~Schuch, D.~Perez-Garcia, and I.~Cirac, Phys.~Rev.~B {\bf 84}, 165139 (2011).
\bibitem{Wen1dfull} X.~Chen, Z.-C.~ Gu and X.-G.~Wen, Phys. Rev. B {\bf 84}, 235128 (2011).
\bibitem{WenCohoBoson}X.~Chen, Z.-C.~Gu, Z.-X.~Liu, and X.-G.~Wen, Phys. Rev. B {\bf 87}, 155114 (2013).
\bibitem{Ashvin2d} Y.-M.~Lu and A.~Vishwanath, Phys. Rev. B {\bf 86}, 125119 (2012).
\bibitem{VS} A.~Vishwanath and T.~Senthil, Phys. Rev. X {\bf 3}, 011016 (2013).

\bibitem{Hermele} A.~M.~Essin and M.~Hermele, Phys. Rev. B {\bf 87}, 104406 (2013). 
\bibitem{Ran} A.~Mesaros and Y.~Ran, Phys. Rev. B {\bf 87}, 155115 (2013).
\bibitem{WenSET} L.-Y.~Hung, X.-G.~Wen, Phys. Rev. B {\bf 87}, 165107 (2013).
\bibitem{AshvinZ2Z2} Y.-M.~Lu and A.~Vishwanath, arXiv:1302.2634.
\bibitem{Chong} C.~Wang and T.~Senthil, 	arXiv:1302.6234.
\bibitem{AshvinToricf} F.~J.~Burnell, X.~Chen, L.~Fidkowski and A.~Vishwanath, arXiv:1302.7072.
\bibitem{LukaszTSC}L.~Fidkowski, X.~Chen and A.~Vishwanath, 	arXiv:1305.5851. 


\bibitem{Redlich} A.~N.~Redlich, 
Phys.~Rev.~Lett. {\bf 52}, 18 (1984);~
Phys. Rev. D {\bf 29}, 2366 (1984).

\bibitem{Semenoff} A.~J.~Niemi and G.~W.~Semenoff, Phys. Rev. Lett. {\bf 51}, 2077 (1983).

\bibitem{FionaMike} M.~Mulligan and F.~J.~Burnell, arXiv:1301.4230.

\bibitem{StatWitt} M.~A.~Metlitski, C.~L.~Kane and M.~P.~A.~Fisher, 	arXiv:1302.6535.


\bibitem{FuKaneMajorana} L.~Fu and C.~L.~Kane, Phys.\ Rev.\ Lett.\ {\bf 100}, 096407 (2008).

\bibitem{Balian} R.~Balian and N.~R.~Werthamer, Phys.\ Rev.\, {\bf 131}, 1553 (1963).
\bibitem{Volovik} G.~E.~Volovik, {\it The Universe in a Helium Droplet.}, International Series of Monagraphs
on Physics. OUP Oxford, (2009).

\bibitem{WW} K.~Walker and Z.~Wang, Frontiers of Physics {\bf 7}, 150 (2012).


\bibitem{ReadGreen} N.~Read and D.~Green, Phys. Rev. B {\bf 61}, 10267 (2000).
\bibitem {Ivanov2001} D.~Ivanov, Phys. Rev. Lett. {\bf 86}, 268 (2001).
 \bibitem{Stern04} A.~Stern, F.~von~Oppen and E.~Mariani, Phys. Rev. B {\bf 70}, 205338 (2004).
 
\bibitem{Stone06} M.~Stone and S.-B.~Chung, Phys. Rev. B {\bf 73} (2006).

\bibitem{KitaevHoneycomb} A.~Kitaev, Annals of Physics {\bf 321} (2006). 

\bibitem{Read2008}N.~Read, Phys. Rev. B {\bf 79}, 045308 (2009).


\bibitem{SCZ}X.~L.~Qi, T.~L.~Hughes, and S.-C.~Zhang, Phys. Rev. B {\bf 78}, 195424 (2008); Phys. Rev. B {\bf 81}, 159901(E) (2010).
\bibitem{Witten}E.~Witten, Phys. Lett. B {\bf 86}, 283 (1979).
\bibitem{Franz}G.~Rosenberg and M.~Franz, Phys. Rev. B {\bf 82}, 035105 (2010).

\bibitem{BNQ} P.~Bonderson, C.~Nayak and X.-L.~Qi, to appear.
\bibitem{AshvinETI} X.~Chen, L.~Fidkowski and A.~Vishwanath, to appear.
\bibitem{ChongDrewSurface} C.~Wang, D.~Potter and T.~Senthil, to appear.
\bibitem{ChongDrewBulk} C.~Wang, D.~Potter and T.~Senthil, to appear.

\bibitem{Goldhaber} A. ~Goldhaber, Phys.~Rev.~Lett {\bf 36}, 1122 (1976).



  

\end{thebibliography}
\end{document}